\documentstyle[12pt,a4,psfig,epsfig]{article}
\newcommand{\vs}[1]{\rule[- #1 mm]{0mm}{#1 mm}}

\begin{document}
\renewcommand{\thefootnote}{\fnsymbol{footnote}}
\newpage
\pagestyle{empty}
\setcounter{page}{0}

\null
\begin{center}

{\Large {\bf PHOTOPRODUCTION OFF NUCLEI}}\\[0,5cm]
{\Large {\bf Part II: Particle and Jet Production}}

\vs{5}

{\bf S.\ Roesler\footnote{present address: CERN, CH-1211 Gen\'eve 23,
Switzerland}
}\\
{\em Universit\"at Siegen, Fachbereich Physik, D-57068 Siegen, Germany}

\vs{0,5}

{\bf R.\ Engel\footnote{present address: 
University of Delaware, Bartol Research Institute, Sharp Lab., 
Newark, DE 19716, U.S.A.}
}\\
{\em Universit\"at Leipzig, Fachbereich Physik, D-04109 Leipzig, Germany
\\ and Universit\"at Siegen, Fachbereich Physik, D-57068 Siegen, Germany}

\vs{0,5}

{\em and}

\vs{0,5}

{\bf J.\ Ranft\footnote{present address:
Universit\"at Siegen, Fachbereich Physik, D-57068 Siegen, Germany}
}\\
{\em Departamento de F\'\i sica de Part\'\i culas, Universidade de Santiago
     de Compostela,\\
     E-15706 Santiago de Compostela, Spain}

\end{center}
\vs{5}

\centerline{ {\bf Abstract}}
\indent
High energy multiparticle photoproduction off nuclear targets is studied.
The photon is assumed to interact
in direct scattering processes or as a resolved $q\bar{q}$-state according
to the Generalized Vector Dominance Model. In the description of
resolved interactions multiple soft and hard processes in each
$q\bar{q}$--nucleon interaction are taken into consideration.
The model, formulated within the framework of the two-component Dual Parton 
Model, is shown to agree with hadron production data from real and weakly
virtual photon--nucleus interactions.
Differences between multiparticle production in photon--nucleus and
hadron--nucleus interactions and features of jet production in 
photon--nucleus collisions at HERA-energies are discussed.

\vfill
\rightline{Siegen SI 96-14}
\rightline{Santiago de Compostela US-FT/45-96}
\rightline{revised version, November 1997}
\vfill

\newpage
\pagestyle{plain}
\renewcommand{\thefootnote}{\arabic{footnote}}
%
%
\section {Introduction}
\noindent
Many features of high energy photoproduction off hadrons
are well understood within the QCD-improved parton model: the photon may
couple directly to a parton of the hadron ({\it direct} processes) or it
may enter the scattering process as a hadronic fluctuation
({\it resolved} processes) (recent reviews are given 
in~\cite{Drees93b,Drees95a}). Since the lifetimes of these 
hadronic fluctuations
are typically long enough to develop properties of ordinary hadrons they
can interact in soft and hard resolved processes~\cite{Bauer78,Kolanoski88a}.
The classification into direct and resolved photon interactions has been
confirmed by many experiments~\cite{Tanaka92,Ahmed92a,Derrick94a}.
Moreover, it has been shown that both classes of processes exhibit different 
properties concerning multiparticle 
production~\cite{Bauer78,Kolanoski88a,Tanaka92,Aid95c,Derrick95f}.
Here, we extend a previous study of photon--nucleus cross sections and their 
high energy shadowing behavior~\cite{Engel96g} to particle production.
This is done by applying the two-component
Dual Parton Model (DPM)~\cite{Capella87b,Bopp94a,Aurenche92a,Engel95a,Engel95d}
to photoproduction off nuclei. 

The two-component DPM is based on the DPM describing soft particle 
production (for a review see~\cite{Capella94a}) and treats
high-$p_\perp$ processes using lowest order perturbative QCD.
This model has been proven to be very reliable in describing 
the main features of hadron--hadron~\cite{Bopp94a,Aurenche92a},
photon--hadron~\cite{Engel95a,Engel95d}, photon--photon~\cite{Engel95d}, 
hadron--nucleus~\cite{Ranft95a}, 
and nucleus--nucleus collisions~\cite{Ranft95a}.
Here, the two-component DPM is applied to photon--nucleus collisions
for the first time.
The two channels which are of particular importance as a basis for our 
present study are hadron--nucleus and photon--hadron collisions.

In hadron--nucleus collisions the multiple interaction process of the hadron 
with target nucleons which explains the ``shadowing''-effect, is understood
in the framework of the Gribov--Glauber 
approximation~\cite{Glauber55,Gribov69b-e,Bertocchi72}.
The two-component DPM incorporates the Gribov--Glauber approximation and
treats each hadron--nucleon interaction by pomeron and reggeon 
exchanges. {\sc Dtunuc 2.0}, a Monte Carlo (MC) realization of the 
two-component DPM for hadron--nucleus and nucleus--nucleus collisions, 
includes the Glauber-formalism~\cite{Shmakov89}, the 
{\sc Phojet} model~\cite{Engel95a,Engel95d} for 
each hadron--nucleon or nucleon--nucleon collision, a formation zone
intranuclear cascade model, and models for spectator deexcitation and
disintegration~\cite{Ferrari95a,Ferrari96a}. As it will be shown further 
below, results on particle production in hadron--nucleus collisions obtained 
with {\sc Dtunuc 2.0} are in good agreement with data.

The description of photon--hadron interactions within the two-component DPM
takes the above mentioned dual nature of the photon into consideration.
In analogy to hadron--hadron interactions, the resolved photon interactions 
are described in terms of multiple soft and hard pomeron and soft reggeon 
exchanges. Parton jet and minijet production in resolved processes as well 
as direct processes are calculated in lowest order perturbative QCD. 
The MC realization {\sc Phojet} has been 
compared in several applications to photoproduction data from the HERA 
collider and a reasonable agreement was found~\cite{Aid95c,Aid95b,Aid96a}.

Using all informations from hadron--nucleus and photon--hadron collisions
it can be expected that the two-component DPM also gives a 
reasonable description of multiparticle photoproduction
off nuclei up to very high energies. 
Such a model might be useful for the study of general properties of
photon--nucleus interactions as well as for the calculation of theoretical
predictions (for example at HERA energies), cosmic ray cascade simulations,
or shielding problems at TeV-lepton colliders.

In Sect.~2 we summarize the basic ideas of the description of hadron--nucleon
and photon--nucleon interactions within the two-component DPM.
Furthermore, its application to photon--nucleus interactions is presented.
In Sect.~3 we demonstrate that the model is able to describe characteristic
features of measured particle production in photon--nucleus interactions. 
The nuclear dependence of multiparticle photoproduction off nuclei and 
direct as well as resolved jet production
are studied. Predictions for HERA-energies are given. 
Finally, in Sect.~4 we summarize our results.
%
%
\section{The two-component Dual Parton Model}
\subsection{\label{phojet} Photon--nucleon interactions}
In the following, we give a summary of the basic ideas of the 
two-component DPM with emphasis to the description of photon--nucleon 
interactions as implemented in the MC generator {\sc Phojet}. More detailed
discussions can be found in Refs.~\cite{Engel95a,Engel95d}.

The physical photon state is approximated as a superposition of a bare
photon and of virtual hadronic states having the same quantum numbers 
as the photon. The bare photon may interact in direct processes.
This direct contribution is
estimated by lowest order perturbative QCD.
The interactions of the hadronic fluctuations of the photon are
called resolved processes and are described within the framework 
of the two-component DPM in terms of reggeon and pomeron exchanges.
We distinguish soft and hard resolved processes according to the transverse
momenta of the intermediate states of the pomeron exchange graph:
By definition, scattering processes resulting in
at least one parton having a transverse momentum larger than a 
cutoff momentum $p_\perp^{\mbox{\scriptsize cutoff}}=3$~GeV/$c$ are called hard 
interactions. All other processes are classified as soft interactions.

The amplitude describing the scattering of the hadronic fluctuation $V$
with a nucleon $N$ is parametrized using a two-channel eikonal
formalism~\cite{Engel95a,Aurenche92a}. In impact parameter representation, 
the amplitude reads~\cite{Engel95a}
\begin{equation}
a_{VN}(s,b) = \frac{i}{2} \left( 1 - e^{-\chi(s,b)}\right),
\label{eff-amp}
\end{equation}
with the eikonal function
\begin{equation}
\chi (s,b)=\chi_{\mbox{\scriptsize S}}(s,b)+\chi_{\mbox{\scriptsize H}}(s,b)+
           \chi_{\mbox{\scriptsize D}}(s,b)+\chi_{\mbox{\scriptsize C}}(s,b).
\end{equation}
Here, the $\chi_{i}$'s denote the contributions from the different
Born graph amplitudes: soft pomeron and reggeon (S), hard pomeron (H), 
triple- and loop-pomeron (D), and double-pomeron amplitudes (C).
The Born graph amplitudes of the soft processes are calculated using Gribov's
reggeon field theory \cite{Gribov67a-e,Gribov69b-e}. 
Assuming universality of soft
interactions,
the free parameters of the model (pomeron couplings,
pomeron intercept, triple-pomeron coupling, and slope parameters) have been
determined by a global fit to $pp$, $p\bar p$, and $\gamma p$ cross
section and slope data~\cite{Engel95a}. However
this universality cannot be applied to photon interactions involving
large transverse momenta. Lowest order perturbative QCD is used
to calculate the resolved and direct hard photon cross
sections~\cite{Engel95a,Combridge77}.

The optical theorem relates the total cross section (i.e.\ elastic 
scattering, diffraction dissociation, and nondiffractive inelastic scattering)
to the discontinuity of the amplitude (\ref{eff-amp}) taken at vanishing
momentum transfer.
At high energies, the discontinuity can be  expressed as a sum over terms
which correspond to graphs with a certain number of cut pomerons
(Abramovski--Gribov--Kancheli cutting rules~\cite{Abramovski73-e}). Therefore,
one obtains for the exclusive cross section for $k_{c}$ cut soft 
pomerons, $l_{c}$ cut hard pomerons, $m_{c}$ cut triple- and loop-pomeron
graphs, and $n_{c}$ cut double-pomeron graphs
\begin{equation}
\sigma (k_{c},l_{c},m_{c},n_{c},s,b)=
\frac{(2\chi_{\mbox{\scriptsize S}})^{k_{c}}}{k_{c}!}
\frac{(2\chi_{\mbox{\scriptsize H}})^{l_{c}}}{l_{c}!}
\frac{(2\chi_{\mbox{\scriptsize D}})^{m_{c}}}{m_{c}!}
\frac{(2\chi_{\mbox{\scriptsize C}})^{n_{c}}}{n_{c}!} e^{-2\chi (s,b)}
\label{cutpro}
\end{equation}
with
\begin{equation}
\int d^2b \sum_{k_{c}+l_{c}+m_{c}+n_{c}=1}^{\infty} \sigma
(k_{c},l_{c},m_{c},n_{c},s,b) =
\sigma^{\mbox{\scriptsize tot}}_{VN}-\sigma^{\mbox{\scriptsize el}}_{VN} .
\end{equation}
Both resolved soft and hard interactions are
unitarized, i.e. a single $q\bar{q}$--nucleon scattering may be built
up of several soft and hard interactions. In contrast to resolved
processes, absorptive corrections to direct photon interactions are
suppressed by one order of the fine structure constant 
$\alpha_{\mbox{\scriptsize em}}$ and are therefore neglected.

In order to relate the cross sections given in Eq.(\ref{cutpro}) to final 
state parton configurations, we use the equivalence between the reggeon and 
pomeron 
exchange amplitudes and certain color flow topologies~\cite{Veneziano76}.
In Fig.\ \ref{colflo} the color flows of single reggeon (a) and single pomeron
exchange processes (c) are shown together with the corresponding cuts.
Whereas a cut reggeon yields one color-field chain (string) 
(Fig.\ \ref{colflo}b), a cut
pomeron results in two strings (Fig.\ \ref{colflo}d) which are assumed to
fragment independently into hadrons. In the large $N_c$ limit of
QCD (with $N_c$ being the number of colors) the same color flow picture of 
the pomeron is also found in hard interactions.

For pomeron cuts involving a hard scattering, the parton kinematics as well as
the flavors and colors are sampled according to the QCD-improved
Parton Model using leading-order matrix elements. We use the GRV 
parametrizations of the parton distribution functions (PDF) of the 
photon~\cite{Gluck92b} and the nucleon~\cite{Gluck92c}. Both, initial and final 
state parton showers are treated. For the latter the Monte 
Carlo (MC) realization as implemented in 
{\sc Jetset}~\cite{Sjostrand86,Sjostrand87a} is applied.

In case of pomeron cuts without large momentum transfer, we assume the 
partonic interpretation of the DPM: hadronic fluctuations of the photon are 
split into a quark-antiquark pair whereas
baryons are approximated by a quark-diquark pair.
The longitudinal momentum fractions of the partons are given by
Regge asymptotics~\cite{Capella80a,Capella80b,Kaidalov82a,Kaidalov82b}.
The valence quark momentum fraction $x$ inside a nucleon is sampled
according to $\rho(x) \sim (1-x)^{1.5}/\sqrt{x}$ whereas the momentum
fraction of a valence quark inside a hadronic fluctuation of a photon is 
obtained from $\rho(x) \sim 1/\sqrt{x (1-x)}$.
For multiple interaction events, the sea quark momenta are sampled using
a $\rho(x) \sim 1/x$ behavior.
The transverse momenta of the soft partons are not predicted by the DPM.
Here, we assume an exponential distribution 
$d^2N_s/dp_\perp^2 \sim \exp(-\beta p_\perp)$.
The energy-dependent slope parameter $\beta$ is obtained requiring
a smooth transition between the transverse momentum distributions of the
soft constituents and the hard scattered partons.

For single diffractive and central diffractive processes,
the parton configurations are generated using the
ideas described above applied to  pomeron--photon/hadron/pomeron
scattering processes~\cite{Engel96e}.
Hence, a diffractive triple-pomeron or loop-pomeron
cut can also involve hard scattering subprocesses resulting in a
rapidity gap event with jets.

All strings are hadronized using 
{\sc Jetset}~7.3~\cite{Sjostrand86,Sjostrand87a}.

Finally it should be mentioned that the model is limited to
photon virtualities $Q^2\ll s$ with $Q^2<9$~GeV$^2$~\cite{Engel95d}.
\subsection{The event generator DTUNUC~2.0 for photon--nucleus collisions}
The MC event generator {\sc Dtunuc~2.0} for hadron--,
photon--, and nucleus--nucleus collisions is based on its
previous version~\cite{Moehring91,Ranft94a} incorporating
the following new features:
\begin{itemize}
\item {\sc Dtunuc} is extended to the description of photoproduction off 
      nuclei.
\item Particle production in each photon--, hadron--, and nucleon--nucleon 
      interaction is based on the MC realization 
      {\sc Phojet}~\cite{Engel95a,Engel95d} of the two-component DPM 
      which includes multiple soft and hard pomeron exchanges and 
      diffraction.
\item A formation zone intranuclear cascade is treated in both, the
      projectile and target spectators including evaporation of nucleons
      and light nuclei from excited spectators, spectator deexcitation via 
      photon emission, and fragmentation of light spectator 
      nuclei~\cite{Ferrari95a,Ferrari96a}.
\end{itemize}

The generation of a photon--nucleus\footnote{Here, we restrict our 
discussion to photon projectiles. The treatment of hadron--nucleus
collisions is similar; the discussion of nucleus--nucleus collisions is not
within the scope of this paper} particle production event proceeds as 
follows:
The treatment of any interaction starts with sampling of an initial spatial
configuration according to nuclear densities
and of the Fermi-momenta of the nucleons. The number of nucleons involved 
in the scattering process with the photon is obtained according to the
Glauber approximation. We use the MC algorithm of~\cite{Shmakov89}, 
here extended to photon projectiles.
The inelastic cross section for the scattering of a photon with virtuality
$Q^2$ with a nucleus $A$ at a squared photon--nucleon c.m.\ energy $s$ and
at impact parameter $b$ reads~\cite{Engel96g}
\begin{eqnarray}
\label{GVDM-sig_gA}
&&\sigma_{\gamma A}(s,Q^2,b)=4\pi\alpha_{\mbox{\scriptsize em}} 
\int_{M_0^2}^{M_1^2} dM^2\ D(M^2) \left(\frac{M^2}{M^2+Q^2}\right)^2 
\left(1+\epsilon\frac{Q^2}{M^2}\right) \sigma_{VA}(s,Q^2,M^2,b) \\
\label{sig_VA}
&&\sigma_{VA}(s,Q^2,M^2,b)=\int \prod_{j=1}^{A} d^3r_j\
\rho_A(\vec{r}_j) \left(1-\left|
\prod_{i=1}^{A}\left[1-\Gamma(s,Q^2,M^2,\vec{b}_i)\right]\right|^2 \right).
\end{eqnarray}
Corresponding to the assumptions of the Generalized Vector Dominance
Model (GVDM) (see for example~\cite{Bauer78,Donnachie78-a} and references
therein) we integrate in (\ref{GVDM-sig_gA}) over the masses $M$ of the 
hadronic fluctuations of the photon. $D(M^2)$ denotes the density of
hadronic states per unit mass-squared interval.
$\epsilon$ is the ratio between the fluxes of longitudinally and
transversally polarized photons.
Eq.(\ref{sig_VA}) relates the inelastic cross section $\sigma_{VA}$ 
for the interaction
of the hadronic fluctuation with the nucleus to the
amplitude $\Gamma$ for the interaction of the hadronic fluctuation with
a nucleon 
\begin{eqnarray}
\label{gamma}
&&\Gamma(s,Q^2,M^2,\vec{b})=
\frac{\sigma_{VN}(s,Q^2,M^2)}{4\pi B(s,Q^2,M^2)}\left(1-i\rho\right)
\exp{\left(\frac{-\vec{b}^2}{2B(s,Q^2,M^2)}\right)}, \\
\label{GVDM-sig_gn-sig_VN}
&&\sigma_{VN}(s,Q^2,M^2)=\frac{\tilde{\sigma}_{VN}(s,Q^2)}
{M^2+Q^2+C^2}\; .
\end{eqnarray}
We refer to~\cite{Engel96g} for the calculation of the $M^2$-independent
quantity $\tilde{\sigma}_{VN}$, the slope $B$, and the parameters entering
the expressions. The
integration over the coordinate space in (\ref{sig_VA}) is performed using 
one-particle Woods--Saxon density distributions $\rho_A$~\cite{Segre77}.
With the impact parameter $b$ and the mass $M$
of the hadronic fluctuation which are sampled according to 
Eqs.(\ref{GVDM-sig_gA},\ref{sig_VA}), we obtain for the probabilities $p_i$
that an inelastic interaction between the projectile and nucleon $i$ 
takes place~\cite{Engel96g,Shmakov89}
\begin{equation}
\label{Gamma}
p_i=\Gamma(\vec{b}_i)+\Gamma^{\star}(\vec{b}_i)
-\Gamma(\vec{b}_i)\Gamma^{\star}(\vec{b}_i)\; .
\end{equation}
Note that $b_i$ is the impact parameter for the interaction of the
hadronic fluctuation with the nucleon $i$ for a fixed spatial configuration 
of nucleons.

As it has been discussed in Sect.~3 of Ref.~\cite{Engel96g}, 
hadronic fluctuations with
$M^2>(2p_{\perp}^{\mbox{\scriptsize cutoff}})^2$ predominantly interact 
in point-like interactions\footnote{Note that in our definition of
``point-like photon interactions''~\cite{Engel96g} interactions of
low-mass $q\bar q$-fluctuations with only one target nucleon are
not included}, i.e. in direct interactions and in 
interactions which are characterized by the anomalous component of the 
photon PDF. In both cases, the virtuality of the $q\bar q$-system
allows to calculate the photon--$q\bar q$ coupling perturbatively.
We denote the corresponding cross sections with 
$\sigma_{\gamma N}^{\mbox{\scriptsize dir}}$ and 
$\sigma_{\gamma N}^{\mbox{\scriptsize ano}}$ and refer to~\cite{Engel96g}
for their calculation within lowest order perturbative QCD. Here we shall
again consider the extreme case that in these processes the hadronic 
fluctuation of the photon interacts
with only one target nucleon. Therefore, in the fraction 
$A\cdot (\sigma_{\gamma N}^{\mbox{\scriptsize dir}}+
\sigma_{\gamma N}^{\mbox{\scriptsize ano}})/ 
\sigma_{\gamma A}^{\mbox{\scriptsize tot}}$ of all events only one hard
resolved (representing the anomalous component) or one direct photon
interaction is sampled.

Furthermore,
as for the calculation of cross sections~\cite{Engel96g}, the coherence 
length of the photon is taken into consideration which effectively leads
to a suppression of the Glauber-cascade at low energies.
Particle production in each inelastic interaction of the photon or its
hadronic fluctuation with a nucleon
is treated by the MC realization {\sc Phojet} of the two-component DPM
(see Sect.\ref{phojet}).
The photon--nucleon interactions are followed by a formation zone 
intranuclear cascade in the target spectator and by subsequent 
evaporation processes of nucleons and light nuclei as well as by 
spectator deexcitation, and by the fragmentation of light spectator
nuclei~\cite{Ferrari95a,Ferrari96a}.
%
%
\section{Particle production}
For studying photoproduction off nuclei we assume the logical sequence to 
be as follows: As mentioned earlier soft hadronic interactions exhibit
universal features irrespective of the nature of the colliding particles.
This fact was already emphasized by Engel and Ranft in~\cite{Engel95d}
where the description of high energy hadron--hadron and photon--hadron
interactions based on the two-component DPM was extended to photon--photon 
interactions. Here, we want to proceed in an analogous way. 
We start from the description of photon--hadron and hadron--nucleus
interactions and show that the model may also provide a reasonable 
description of photon--nucleus interactions.

Photon--hadron collisions have been discussed in the framework of the
two-component DPM elsewhere~\cite{Engel95a,Engel95d}. Furthermore, also
numerous studies of hadron--nucleus collisions within this model 
exist (see for instance~\cite{Moehring91,Ranft94a,Ranft95a} and references
therein). Of course for the latter interaction channel it has to be 
verified that the new version of {\sc Dtunuc} is able to describe
the data. Since this has not yet been demonstrated we give
a few examples further below.

Based on the reasonable description of these two interaction channels
the model is then applied to multiparticle photoproduction off nuclei. This
application may serve as a severe test of the model since it carries no 
further freedom.
\subsection{Hadron--nucleus interactions}
In Table~\ref{hAmulttab} average shower particle multiplicities calculated
for interactions of positively charged pions and protons with different
target nuclei at 50, 100, and 150~GeV are given. Here, shower particles are
defined as singly charged particles with Lorentz-$\beta$ values exceeding
0.7. These results are compared to data taken from Ref.~\cite{Braune83}.
For all configurations there is a reasonable agreement between the
calculated and measured values.

The distribution of the multiplicity of shower particles produced
in interactions of 525~GeV pions in nuclear emulsions is plotted in 
Fig.\ \ref{pAmul}a. Model results shown as histogram are compared to 
data~\cite{Cherry94}. As this plot demonstrates also the distribution of the 
shower particle multiplicity is reproduced by the model.

Calculated and measured~\cite{Cherry94} pseudorapidity distributions of shower 
particles are shown in Fig.\ \ref{pAmul}b again for pion--emulsion interactions
at 525~GeV and, additionally, for 60~GeV.

The invariant $\pi^0$ cross section as a function of the transverse
momentum in proton--gold collisions at 200~GeV has been measured by the
WA80-Collab.~\cite{Albrecht90}. In Fig.\ \ref{pApt} we compare our 
calculations to these data and find a reasonable agreement. 
Corresponding to the kinematic cuts of the experiment the pseudorapidity 
range $1.5\le \eta \le 2.1$ is considered only. 
\subsection{Photon--nucleus interactions}
\subsubsection{General properties of the model}
Before comparing model results to data from lepton--nucleus interactions 
let us first outline differences between hadron-- and photon--nucleus
collisions with respect to properties of the Glauber-cascade and 
multiplicities at fixed projectile energy. In lepton--nucleus interactions
the projectile photon has varying energies and virtualities and a direct
comparison with hadron--nucleus collisions would be less conclusive.

In Fig.\ \ref{iAmult} we present average numbers of target nucleons 
interacting with the projectile $\langle \nu_t\rangle$ and multiplicities 
of shower $\langle N_s\rangle$ and heavy particles $\langle N_h\rangle$ 
(charged particles with $\beta\le 0.7$) calculated for
pion--, real photon--, and weakly virtual photon--copper
interactions at energies of the projectile in the nucleus rest system
between 10~GeV and 100~TeV. Differences between the $\nu_t$-values 
for pion and photon projectiles at fixed energy arise from differences
between the pion--nucleon cross section and the (averaged over all masses $M$) 
$q\bar q$--nucleon cross section $\sigma_{VN}$ (Eq.(\ref{GVDM-sig_gn-sig_VN}))
which enter the calculations~\cite{Engel96g}. 
Since the latter is slightly smaller,
the real photon interacts with less nucleons than the pion.
%
%

Furthermore, $\langle \nu_t \rangle$ decreases with the photon virtuality
at fixed energy. There are two main effects being responsible for this
behavior:
\begin{itemize}
\item[(i)] Due to the $1/(M^2+Q^2)$-behavior of  $\sigma_{VN}$ this
           cross section decreases with $Q^2$ at fixed $M^2$ and 
           fixed energy. Moreover,
           at large photon virtualities interactions of resolved photons 
           with large masses $M^2$ become more important leading to a further
           decrease of the cross section $\sigma_{VN}$ and therefore
           $\langle \nu_t \rangle$.
\item[(ii)]The fraction of events with point-like photon interactions 
           ($\nu_t=1$) increases with rising photon virtuality.
\end{itemize}
As an example, this is shown in 
Fig.\ \ref{icunutnh}a for photon--copper interactions at a photon--nucleon c.m.\
energy of 150~GeV, an energy 
which could be available in the future for photon--nucleus collisions at 
HERA~\cite{Arneodo96}. In our model, 
the average heavy particle multiplicity is approximately proportional
to $\nu_t$ irrespectively of the nature of the projectile.
This dependence is shown in Fig.\ \ref{icunutnh}b for protons, pions,
real and weakly virtual photons ($Q^2=2$~GeV$^2$). It reflects the fact
that the number of heavy particles depends only on the number of nucleons
``knocked out'' of the target by the projectile and by subsequent
intranuclear cascade processes and not on particular properties of the
projectile--nucleon interactions~\cite{Ferrari95a,Ferrari96a}.

\subsubsection{\label{ga_avmul}Average multiplicities}
Average shower particle multiplicities in interactions of 150~GeV muons in
emulsions are compared to data~\cite{Hand78} in Fig.\ \ref{mupemul150nom}.
Corresponding to the measurements, from the calculated events only those
with more than two heavy particles in the final state have been taken into
consideration. The flux of virtual photons is sampled
according to the EPA and to the $Q^2$-dependence of the photon--nucleus
cross sections~\cite{Engel96g}.
The multiplicities are shown as function of the inverse of the Bjorken-$x$.
In this representation the average photon virtuality is decreasing
from about 8~GeV$^2$ in the lowest bin ($1/x_{\mbox{\scriptsize Bj}}<10$)
to about 0.9~GeV$^2$ in the highest bin~\cite{Hand78}. 
Our results agree with the
data in the kinematic region of photoproduction ($Q^2\stackrel{<}{\sim} 
4$~GeV$^2$, $1/x_{\mbox{\scriptsize Bj}}>25$) but underestimate the
data for higher virtualities. This fact may indicate that the 
{\sc Phojet} realization of the two-component DPM for photon--nucleon 
interactions fails in describing particle production in interactions
of photons with relatively high virtualities at low energies.
However, whereas the calculations are based in total on 50000 events one has 
to note that the data include
about 17 events in each $1/x_{\mbox{\scriptsize Bj}}$-bin~\cite{Hand78}.
Taking into account that these 17 events involve interactions with six
different targets (emulsion components) at different collision energies
and photon virtualities one might conclude that the statistical 
significance of the data is limited.

Finally, we compare our results on multiplicities in muon--deuterium and
muon--xenon interactions to data of the E665-Collab.~\cite{Adams94b}.
The experiments were performed with a 490~GeV positive muon beam.
The kinematic region under investigation is $Q^2>1$~GeV$^2$, 
$8<W<30$~GeV, $x_{\mbox{\scriptsize Bj}}>0.002$, and $0.1<\nu/E_\mu<0.85$
(with $\nu$ being the photon energy in the laboratory). 
Furthermore, only charged particles with momenta $p>200$~MeV/$c$ in the 
laboratory frame are considered.
We note that although the distribution of photons in leptons decreases
with increasing $Q^2$ still a considerable fraction of all events is
characterized by photons with rather large virtualities. These events
cannot be expected to be described reliably within the present approach.
In Figs.\ \ref{mupA490muhpm}a and \ref{mupA490muhpm}c 
the energy dependence of the average total, positively and negatively 
charged hadron multiplicities are 
shown. For both target nuclei the data are well reproduced by the model.
The multiplicities of charged hadrons are shown separately for the forward 
and backward 
region of the photon--nucleon c.m.\ frame in Figs.\ \ref{mupA490muhpm}b and
\ref{mupA490muhpm}d.
The calculated multiplicities in the backward region which are strongly
affected by target associated particle production and, therefore, by the
laboratory momentum cut applied to the final state hadrons, are slightly
higher than the data whereas those in forward direction are lower.
Multiplicities averaged over all energies of the photon--nucleus interaction 
are compared to the corresponding experimental values in 
Table~\ref{mupAmultab}.
\subsubsection{Inclusive particle distributions}
In Fig.\ \ref{iemuletans} the model results for the pseudorapidity 
distributions
of shower particles from muon--emulsion interactions at 150~GeV are
compared to data again from measurements by Hand {\it et al.}~\cite{Hand78}.
The kinematic range of the experiment is $0.6<Q^2<21$~GeV$^2$ and 
$2.5<W\le 16.5$~GeV. The distributions shown for muon--emulsion interaction
cover different but overlapping $W$-ranges: in a) data and MC results are 
plotted for $9<W<14$~GeV and in b) for $W>10$~GeV 
($\langle Q^2\rangle=2.7$~GeV$^2$).
Let us first compare the two data sets which include 43 events in a) and 47
events in b)~\cite{Hand78}. 
One obvious difference is the peak in the distribution in a)
at $\eta=1.25$ which is not present in b). 
The only energy range which is not covered by the data in Fig.\ 
\ref{iemuletans}a as compared to \ref{iemuletans}b is $9<W<10$~GeV.
Therefore, we assume that
this peak is due to statistical uncertainties within the experiment 
(c.f. also our discussion in Sect.\ref{ga_avmul}).
Comparing the model results to the data we note that agreement within
the statistical errors is obtained for $\eta>2$. Disregarding
the above mentioned peak, at lower pseudorapidities we underestimate the 
measured distributions slightly. Furthermore the comparison suggests that
both, the measured and calculated distributions, agree in normalization
(i.e. in the average multiplicity of shower particles) but the maximum of
the calculated distribution appears at somewhat higher pseudorapidities
than the measured one.
In order to understand this discrepancy we compare in Fig.\ \ref{iemuletans}a
in addition results for pion--emulsion interactions at 60~GeV to 
data~\protect\cite{Babecki78}. This energy corresponds to the average
photon laboratory energy of the distribution for the muon projectile in 
this Figure. The cuts on both data sample are similar:
only events with at least three heavy particles are considered and 
shower particles are defined by their Lorentz-$\beta$ value as $\beta>0.7$.
Our results for $\pi^-$--emulsion interactions agree in shape and in position 
of the maximum to the data.
Therefore and with respect to conclusions from comparisons with data
drawn further below we attribute the discrepancy in the position of the 
maximum to the statistical uncertainties of the experiment.

Turning again to muon--deuterium and muon--xenon interactions at 490~GeV we
compare in Figs.\ \ref{mupdeu490yyhpm} and \ref{mupxe490yyhpm} rapidity
distributions of positively (a) and negatively charged hadrons (b) in the 
photon--nucleon c.m.\ frame to data~\cite{Adams94b}. The comparisons are
shown for three ranges of the photon--nucleon c.m.\ energy $W$. Taking into
account that the treatment 
of high $Q^2$-values in our model~\cite{Engel95d} might be too simplified, 
the description of the data by the model is satisfactory with the
exception of the production of positively charged particles in the
target fragmentation region of the muon--xenon interaction 
(Fig.\ \ref{mupxe490yyhpm}a, $y\approx -3$). The peak in the calculated
distributions clearly reflects the production of target associated
particles by intranuclear cascade processes which are present in the
distributions if a momentum cut as low as 200~MeV/$c$ in the laboratory frame 
is applied to the results. The multiplicity seen in the target fragmentation
region depends strongly on this cutoff.
We assume that the differences might be due to 
additional kinematic cuts applied to the data or due to experimental
uncertainties (with respect to the momentum cutoff) for the following reasons: 
(i) the models for slow particle production 
implemented in {\sc Dtunuc 2.0} are in good agreement with data from
hadron--nucleus as well as nucleus--nucleus 
collisions~\cite{Ferrari95a,Ferrari96a}, (ii) the dependence of shadowing 
on the photon virtuality and energy is qualitatively understood within our 
model and describes corresponding data of the E665-Collab. reasonably 
well~\cite{Engel96g}, and (iii) the rapidity distributions of negatively 
charged particles and of positively charged particles outside the target 
fragmentation region in muon--xenon interactions and of charged particles in 
muon--deuterium interactions agree with E665-data.

Distributions of charged hadrons from muon--deuterium and muon--xenon
interactions at 490~GeV were measured as function of $z=E/\nu$ ($E$ and
$\nu$ being the secondary hadron energy and the photon energy in
the target rest frame, resp.) 
by Adams {\it et al.}~\cite{Adams94a}.
In Figs.\ \ref{mupA490zhpm}a and \ref{mupA490zhpm}b our results on
the $z$-distributions are shown together with these data. 
Both, model results and data, are restricted to the shadowing region, i.e. to
$x_{\mbox{\scriptsize Bj}}<0.005$ and $Q^2<1$~GeV$^2$.
We find a good agreement in the whole $z$-range.

Feynman-$x$ distributions are usually studied in
terms of the one-particle inclusive variable $F(x_{\mbox{\scriptsize F}})$
defined as
\begin{equation}
\label{xf_def}
F(x_{\mbox{\scriptsize F}})=\frac{1}{\sigma_{\gamma A}^{\mbox{\scriptsize tot}}}
\frac{2E}{\pi W}\frac{d\sigma}{dx_{\mbox{\scriptsize F}}}, \qquad
x_{\mbox{\scriptsize F}}=\frac{2p_\parallel}{W}.
\end{equation}
The quantities $E$, $p_\parallel$, and $W=\sqrt{s}$ denote the energy and
longitudinal momentum in the photon--nucleon c.m.\ system, and the 
photon--nucleon c.m.\ energy, resp.
The Feynman-$x$ distribution of positively
and negatively charged hadrons is given together with data~\cite{Loomis79}
in Fig.\ \ref{mupA490zhpm}c for muon--deuterium interactions at 147~GeV.
Here, the photon virtualities are restricted to the range 
$0.5<Q^2<3$~GeV$^2$ and the photon--nucleon c.m.\ energies to $W>10$~GeV.

Finally, we compare in Fig.\ \ref{mupdeu147pt}a the transverse momentum
distributions of charged particles in different ranges of 
$x_{\mbox{\scriptsize F}}$ again to data on muon--deuterium interactions from 
Ref.~\cite{Loomis79}. Our results are presented as histograms whereas
fits to measured $p_\perp$-distributions are shown as continuous lines.
The experimental uncertainties are increasing with $p_\perp$ and are at
least of the order of the differences between the fits for positive and
negative particles~\cite{Loomis79}. They do not allow to draw conclusions
concerning the disagreement between model predictions and data at large
transverse momenta.
The dependence of the average transverse momenta of charged particles
on $x_{\mbox{\scriptsize F}}$ is shown in Fig.\ \ref{mupdeu147pt}b. 
\subsubsection{The nuclear dependence of particle production}
In order to study the dependence of particle production on the mass number
$A$ of the nuclear target, inclusive single particle cross sections, like
pseudorapidity ($d\sigma/d\eta$), transverse
momentum ($d\sigma/dp_{\perp}^2$), or Feynman-$x$
distributions ($d\sigma/dx_{\mbox{\scriptsize F}}$), are usually fitted
to a $A^{\alpha}$-behavior.
In the projectile fragmentation region of photon--nucleus collisions we
expect $\alpha$ to approach unity (i) for small $q\bar{q}$--nucleon cross
sections, such as at low energies or for $Q^2>0$, (ii) at high energies
for interactions becoming more point-like, and (iii) at large transverse
momenta due to hard interactions.
In the fragmentation region of the target nucleus $\alpha$ can exceed the
value of one due to the formation zone intranuclear cascade, an effect being
outside the scope of the present paper. Here, we restrict our discussion to 
the photon fragmentation region.

In Fig.\ \ref{iAalp} we show fits to results of the model on interactions
of real and weakly virtual photons ($Q^2=1$~GeV$^2$) with carbon,
aluminum, copper, silver, xenon, and gold. Only charged particles with
Lorentz-$\beta>0.7$ (shower particles) are taken into consideration.
The dependences of $\alpha$ on the pseudorapidity (a), on the transverse
momentum (b), and on the Feynman-$x$ 
variable (c) are plotted for a laboratory energy of 250~GeV.
At all values of the studied variables $\alpha$ is larger for virtual
than for real photon projectiles. This stronger $A$-dependence of the cross 
sections for virtual photons is due to the $Q^2$-dependence of the effective
$q\bar q$--nucleon cross sections $\sigma_{VN}$ 
(Eq.(\ref{GVDM-sig_gn-sig_VN})). It demonstrates that the previously discussed
property of the model, the
$Q^2$-dependence of shadowing (see Fig.\ \ref{icunutnh}), is clearly
visible also in inclusive particle distributions.
It might be interesting to study this effect also experimentally.
\subsection{Jet photoproduction off nuclei at HERA-energies}
Studying jet
production in direct and resolved photon events in $\gamma p$ interactions
striking differences have been observed by the 
experiments~\cite{Tanaka92,Aid95c,Derrick95f}. Therefore one
can expect to observe similar effects when comparing point-like and
resolved photon processes in photon--nucleus interactions.
In the following, properties of hadronic jets produced in interactions of
real photons with nuclei are studied at energies which might be reached
with nuclear beams at HERA. Assuming nuclei with equal numbers of protons 
and neutrons the nuclear beam will have an energy per nucleon of 
410~GeV~\cite{Arneodo96}.
With an average photon energy of about half of the electron energy 
(27.6~GeV) the photon--nucleon c.m.\ energy will be about 150~GeV.

Before discussing jet production let us first compare transverse 
momentum distributions of charged particles in proton-- and photon--nucleus
collisions. They very clearly reflect differences between interactions with
proton and the photon projectiles which are responsible for characteristic
features of jet production in photon--nucleus collisions.
As shown in Fig.\ \ref{iiherapt} the $p_\perp$-distribution of charged 
particles for 
photon-projectiles exhibits a less rapid decrease and extends to higher
transverse momenta than the one for hadron projectiles.
This property was already discussed in~\cite{Engel95d} comparing 
hadron--hadron, photon--hadron, and photon--photon interactions. The reasons
are (i) the photon may interact in direct processes, and (ii) the photon PDFs 
are ``harder'' as compared to the proton PDFs.
 
As it has been shown in the analysis of data from HERA on photon--proton 
collisions~\cite{Erdmann96} the study of jet production may give evidence
for the subdivision into direct and resolved photon interactions. 
Therefore it can be expected that differences between hadron--nucleus
and photon--nucleus collisions and characteristic features of point-like
photon interactions emerge most clearly in a study of jet production.
In the following, all observables used refer to the photon--nucleon c.m.\
system. Particle jets are defined according to the 
Snowmass-convention~\cite{snowmass-jets}.
In the plane of pseudorapidity $\eta$ and azimuthal angle $\phi$
a jet is defined as a collection of particles contained in
a cone of radius $R=\sqrt{(\Delta\eta)^2+(\Delta\phi)^2}=1$.
The jet transverse energy $E_\perp^{\mbox{\scriptsize jet}}$ is taken as the
sum of the particle transverse energies inside the cone.
The jet pseudorapidity $\eta^{\mbox{\scriptsize jet}}$ is calculated as
$E_\perp$-weighted average over the pseudorapidities of the particles
belonging to the jet.

In Fig.\ \ref{iAjet}a we compare the transverse energy distributions of
jets from proton--carbon and photon--carbon interactions.
Similar to proton/photon--proton interactions~\cite{Engel95d}, 
the $E_\perp^{\mbox{\scriptsize 
jet}}$ distribution extends to higher transverse energies  for photon--nucleus 
collisions as compared to proton--nucleus collisions. Differences between
the two channels are even more pronounced within the pseudorapidity
distributions of the jet axes. This is shown in Fig.\ \ref{iAjet}b for
jets with transverse energies higher than 6~GeV and 10~GeV.
As initially mentioned, since the PDFs of the photon are harder
than the PDFs of the proton and due to the point-like photon interactions
we obtain in photon--nucleus collisions considerably more jets in 
forward direction ($\eta^{\mbox{\scriptsize jet}}>0$). In backward direction 
the soft underlying event, being more pronounced for proton
projectiles, is responsible for a higher jet rate.

The kinematic regions of jets from direct and resolved photon interactions
might be separated if jet production is analysed as function of
\begin{equation}
x^{\mbox{\scriptsize obs}}=\frac{
E_\perp^{\mbox{\scriptsize jet1}}e^{\eta^{\mbox{\tiny jet1}}}+
E_\perp^{\mbox{\scriptsize jet2}}e^{\eta^{\mbox{\tiny jet2}}}}
{2E_\gamma}\; .
\end{equation}
$E_\gamma$ denotes the photon energy.
$x^{\mbox{\scriptsize obs}}$ can be considered as an estimate for the
fraction of the projectile momentum entering the hard 
scattering~\cite{Aid95c,Ahmed95e}. In Fig.\ \ref{gAjtxg}a
we show $x^{\mbox{\scriptsize obs}}$-distributions for proton--carbon,
photon--carbon, and photon--sulfur interactions.
Point-like photon interactions contribute exclusively to the
region $x^{\mbox{\scriptsize obs}}>0.6$. The specific properties of jets
and of the underlying event are usually studied in terms of jet profiles.
This is shown for photon--carbon interactions and different 
$x^{\mbox{\scriptsize obs}}$-bins in Fig.\ \ref{gAjtxg}b where we plot the 
average transverse energy as a function of the distance in pseudorapidity 
from the jet axis. Since in direct photon interactions there is no photon 
remnant which could contribute to the jet pedestal, the transverse
energy outside the jets is decreasing with increasing $x^{\mbox{\scriptsize 
obs}}$. Of course, as the mass number of the target nucleus increases
the soft underlying event becomes more pronounced, i.e. the average 
transverse energy of the jet pedestal increases. This is demonstrated
in Fig.\ \ref{gAjtxg}c for four different target nuclei. Here, all jets
with $E_\perp>6$~GeV are included.
%
%
\section{Summary and conclusions}
The two-component Dual Parton Model is applied to multiparticle
photoproduction off nuclei.
By incorporating the {\sc Phojet} event generator for hadron--nucleon and 
photon--nucleon interactions, it is now possible to describe with the
event generator {\sc Dtunuc 2.0} consistently cross sections and particle 
production in high energy hadron--nucleus, real and weakly virtual 
photon--nucleus, and nucleus--nucleus collisions (the latter are not discussed 
in this paper).

In~\cite{Engel96g} it has been shown that the model correctly 
describes the shadowing behavior of high energy photon--nucleus cross 
sections. Based on this fact and on the overall good agreement of model 
results and data
in all main aspects of particle production in photon--hadron and 
hadron--nucleus collisions no further
freedom exists in the model for its application to multiparticle 
photoproduction off nuclei. It is therefore a severe test of the model.
Unfortunately, only a few data are available in the kinematic region
$Q^2\ll s$, $Q^2<9$~GeV$^2$
to which the predictions of {\sc Dtunuc 2.0} can be compared.
As discussed in this paper, model results are qualitatively consistent
with many features of these data.

Within the model the photon is treated (i) as a resolved $q\bar{q}$-state 
interacting with target nucleons according to the GVDM and the 
Gribov--Glauber approximation in multiple soft and hard scattering processes
and (ii) as a point-like object interacting in a single hard scattering with 
one target nucleon (anomalous component of the photon PDF and direct
photon interactions). On the basis of this treatment we obtain the following
results:
\begin{itemize}
\item
As expected from studies of photon--nucleus cross sections,
also particle production off nuclei clearly shows decreasing
shadowing with increasing photon virtualities. 
\item
At energies of present
fixed target experiments, inclusive single particle cross sections become 
proportional to $A^\alpha, \alpha\approx 1$, already at $Q^2>1$~GeV$^2$ due to
decreasing $q\bar{q}$--nucleon cross sections with increasing $Q^2$.
However, more data on particle production in photon--nucleus collisions
would be needed for a detailed investigation of the transition region.
\item
In analogy to observations in photon--proton collisions at HERA we argue
that also in photoproduction off nuclei the dual nature of the photon 
and, therefore, point-like photon interactions
show up most clearly in hadronic jet production. As examples we present
transverse energy and pseudorapidity distributions of jets and we study 
jet profiles in real photon--nucleus collisions at energies which might be
reached with nuclear beams at HERA. The strong dependence of the jet profiles
and of the underlying event on the fraction of the momentum of the photon 
going into jets which was first observed at HERA, can be expected to be
present also in photon--nucleus collisions. 
\end{itemize}
%
%
\section*{Acknowledgements}
Discussions with F.\ W. Bopp are gratefully acknowledged.
One of the authors (J.R.) thanks C.\ Pajares for the hospitality
at the University Santiago de Compostela and he was supported by the
Direccion General de Politicia Cientifica of Spain.
One of the authors (R.E.) was supported by the Deutsche Forschungsgemeinschaft
under contract No. Schi 422/1-2.
%
%
\clearpage
\bibliographystyle{prsty}

%
%
\clearpage
\section*{Tables}
\noindent
\begin{table}[htb]
\caption{\label{hAmulttab}
    Multiplicities of shower particles ($\beta>$0.7) in interactions of
    pions and protons with carbon--, copper--, and lead--nuclei at 50, 100,
    and 150~GeV are compared to measurements~\protect\cite{Braune83}.}
\medskip
\begin{center}
\renewcommand{\arraystretch}{1.5}
\begin{tabular}{|r|cccccc|} \hline
& \multicolumn{2}{c}{$E_{\mbox{\scriptsize Lab}}$=50~GeV}&
\multicolumn{2}{c}{$E_{\mbox{\scriptsize Lab}}$=100~GeV}&
\multicolumn{2}{c|}{$E_{\mbox{\scriptsize Lab}}$=150~GeV} \\
& {\sc Dtunuc 2.0} & Exp.& {\sc Dtunuc 2.0} & Exp.& {\sc Dtunuc 2.0} & Exp.
\\ \hline \hline
 $\pi^+$--C &  7.3& 7.62$\pm$0.14&  9.0& 9.19$\pm$0.17&  9.9& 10.01$\pm$0.18
\\ \hline
 $\pi^+$--Cu&  8.6& 8.81$\pm$0.23& 10.5& 10.41$\pm$0.27& 11.8& 11.57$\pm$0.30
\\ \hline
 $\pi^+$--Pb& 10.1&10.11$\pm$0.38& 12.2& 11.54$\pm$0.44& 14.3& 13.08$\pm$0.50
\\ \hline
 p--C &  7.7& 7.88$\pm$0.15&  9.2& 9.25$\pm$0.18& 10.3& 10.58$\pm$0.19
\\ \hline
 p--Cu&  9.8& 9.52$\pm$0.24& 11.9& 11.40$\pm$0.29& 13.3& 12.93$\pm$0.33
\\ \hline
 p--Pb& 11.1&11.31$\pm$0.43& 13.8&               & 15.9& 14.95$\pm$0.57
\\ \hline
\end{tabular}
\end{center}
\end{table}
\begin{table}[htb]
\caption{\label{mupAmultab}
    Results on average multiplicities of all charged
    $\langle N_{\mbox{\scriptsize ch}}\rangle$, positive
    $\langle N_+\rangle$, negative $\langle N_-\rangle$, forward
    $\langle N_{\mbox{\scriptsize F}}\rangle$, and backward
    $\langle N_{\mbox{\scriptsize B}}\rangle$ hadrons ($p>200$~MeV/$c$)
    for muon--deuterium and muon--xenon interactions at 490~GeV are compared 
    to data of the E665-Collab.~\protect\cite{Adams94b}.}
\medskip
\begin{center}
\renewcommand{\arraystretch}{1.5}
\begin{tabular}{|r|cccc|} \hline
& \multicolumn{2}{c}{$\mu^+$--D$_2$}&
\multicolumn{2}{c|}{$\mu^+$--Xe} \\
& {\sc Dtunuc 2.0} & Exp.& {\sc Dtunuc 2.0} & Exp.
\\ \hline \hline
 $\langle N_{\mbox{\scriptsize ch}}\rangle$ &
 7.7 & 7.83$\pm$0.07&10.8 & 9.99$\pm$0.13
\\ \hline
 $\langle N_+\rangle$ &
 4.1 & 4.16$\pm$0.05& 6.8 & 6.00$\pm$0.11
\\ \hline
 $\langle N_-\rangle$ &
 3.6 & 3.67$\pm$0.05& 4.0 & 3.99$\pm$0.06
\\ \hline
 $\langle N_{\mbox{\scriptsize F}}\rangle$ &
 4.1 & 4.60$\pm$0.05& 4.2 & 4.70$\pm$0.07
\\ \hline
 $\langle N_{\mbox{\scriptsize B}}\rangle$ &
 3.6 & 3.23$\pm$0.04& 6.6 & 5.29$\pm$0.11
\\ \hline
\end{tabular}
\end{center}
\end{table}
%
%
\clearpage
\section*{Figure Captions}
\begin{enumerate}
\item \label{colflo}
      Color flow picture of a single reggeon exchange graph (a) and
      the corresponding unitarity cut (b). In c) and d) the same is
      shown for a single pomeron exchange.
\item \label{pAmul}
      Comparison of model results to data on pion--emulsion 
      interactions~\protect\cite{Cherry94}.
      In a) shower particle multiplicity distributions are shown for
      a laboratory energy of 525~GeV. In b) pseudorapidity distributions 
      of shower particles are given for two different energies:
      60~GeV and 525~GeV.
\item \label{pApt}
      Model predictions on transverse momentum distributions
      of $\pi^0$'s in proton--gold interactions at 200~GeV are compared
      to data of the WA80-Collab.~\protect\cite{Albrecht90}. 
\item \label{iAmult}
      Average numbers of copper nucleons $\langle \nu_t\rangle$
      struck by projectile pions, real and virtual ($Q^2=$2~GeV$^2$) photons
      are given as function of the projectile laboratory energy (a).
      In b) and c) average multiplicities of shower $\langle N_s\rangle$
      and heavy particles $\langle N_h\rangle$ are presented.
      Shower particles are defined as singly
      charged particles with $\beta>0.7$. Heavy particles are
      charged particles with $\beta\le 0.7$ except residual copper nuclei.
\item \label{icunutnh}
      In a) the $Q^2$-dependence of the average number of copper nucleons
      interacting with the photon $\langle \nu_t \rangle$ is given for a 
      photon--nucleon c.m.\ energy of 150~GeV.
      In b) the dependence of the average heavy particle multiplicity on 
      $\nu_t$ is shown for proton, pion, real photon, and weakly virtual 
      photon ($Q^2=2$~GeV$^2$) projectiles at the same energy.
\item \label{mupemul150nom}
      Average numbers of charged hadrons ($\beta>0.7$) in muon--emulsion 
      interaction events at 150~GeV with more than two heavy particles in 
      the final state are given as function of the inverse
      of the Bjorken-$x$ variable and are compared to 
      data~\protect\cite{Hand78}.
\item \label{mupA490muhpm}
      The dependence of average multiplicities of charged hadrons
      on the squared energy of the virtual photon--nucleon c.m.\ system
      is shown for
      muon--deuterium (a,b) and muon--xenon interactions (c,d) at 490~GeV  and
      is compared to measurements of the E665-Collab.~\protect\cite{Adams94b}.
      In a) and c) we give the average multiplicities of all
      $\langle N_{\mbox {\scriptsize ch}} \rangle$, of positively
      $\langle N_+ \rangle$, and of negatively
      $\langle N_- \rangle$ charged hadrons. In b) and d) the average
      multiplicities of all charged hadrons are shown for positive 
      $\langle N_{\mbox {\scriptsize F}} \rangle$ and
      negative $\langle N_{\mbox {\scriptsize B}} \rangle$ c.m.\ rapidities.
\item \label{iemuletans}
      Pseudorapidity distributions of charged hadrons ($\beta>0.7$) from
      muon--emulsion interactions at 150~GeV are shown together with 
      data~\protect\cite{Hand78}. In a) the photon--nucleon c.m.\ energy
      range is restricted to $9<W<14$~GeV. In addition, the pseudorapidity 
      distribution of charged hadrons from pion--emulsion interactions at 
      60~GeV is plotted and compared to data~\protect\cite{Babecki78}. 
      In b) the results for $W>10$~GeV are given.
      In both distributions only events with $N_h\ge 3$ are included.
\item \label{mupdeu490yyhpm}
      Rapidity distributions of positive (a) and negative hadrons (b)
      from muon--deuterium interactions at 490~GeV are shown together with
      data~\protect\cite{Adams94b} for different ranges of the virtual 
      photon--nucleon c.m.\ energy $W$.
\item \label{mupxe490yyhpm}
      As in Fig.\ \ref{mupdeu490yyhpm}, here for muon--xenon interactions.
\item \label{mupA490zhpm}
      Energy- ($z$-) distributions of charged hadrons from muon--deuterium (a) 
      and muon--xenon interactions (b) at 490~GeV are compared to
      measurements of the E665-Collab.~\protect\cite{Adams94a}. 
      Corresponding to the experimental cuts applied to the data
      the MC-results are restricted to $x_{\mbox{\scriptsize Bj}}<0.005$, 
      $Q^2<1$~GeV$^2$, and $\nu>100$~GeV. In c) model results for 
      Feynman-$x$ distributions
      of charged hadrons from muon--deuterium interactions at 
      $E_{\mbox{\scriptsize Lab}}=147$~GeV are compared to 
      data~\protect\cite{Loomis79}.
\item \label{mupdeu147pt}
      In a) transverse momentum distributions of charged hadrons
      from muon--deuterium interactions at 147~GeV are presented 
      for different Feynman-$x$ ranges.
      The model results (histograms) are plotted together with fits to
      data as given in~\protect\cite{Loomis79}.
      In b) the  $x_{\mbox{\scriptsize F}}$-dependence of the average
      transverse momenta of charged hadrons are compared to 
      data~\protect\cite{Loomis79}.
\item \label{iAalp}
      The power $\alpha$ of the nuclear dependence of inclusive
      charged particle cross sections is shown as function of the
      pseudorapidity (a), the transverse momentum (b), and the
      Feynman-$x$ variable (c) for real photon and weakly virtual
      photon projectiles of 250~GeV laboratory energy.
\item \label{iiherapt}
      Transverse momentum distributions of charged hadrons from
      proton-- and photon--carbon interactions at a proton/photon--nucleon
      c.m.\ energy of 150~GeV.
\item \label{iAjet}
      In a) transverse energy distributions are presented 
      for hadronic jets from proton--carbon and real photon--carbon
      interactions at an energy of 150~GeV in the proton/photon--nucleon
      c.m.\ system. In b) the
      pseudorapidity distributions of the jet axes are shown for two 
      lower cuts in transverse energy.
\item \label{gAjtxg} 
      Dependence of jet production on the fraction of the projectile momentum
      carried by the jets ($x^{\mbox{\scriptsize obs}}$) at a
      proton/photon--nucleon c.m.\ energy of 150~GeV:
      In a) the $x^{\mbox{\scriptsize obs}}$-distributions of proton-- and
      real photon--carbon and of real photon--sulfur interactions
      are plotted. Jet profiles
      as function of the distance in pseudorapidity from the jet axis
      are presented in b) for different $x^{\mbox{\scriptsize obs}}$-bins.
      In c) the jet profiles are shown for different target nuclei and
      all $x^{\mbox{\scriptsize obs}}$-values.
\end{enumerate}
\clearpage
\newpage
\pagestyle{empty}
\begin{figure}[htb]
\setlength{\unitlength}{1cm}
\begin{picture}(15,23)(0,0)
\put(3.5,14.0){\epsfig{figure=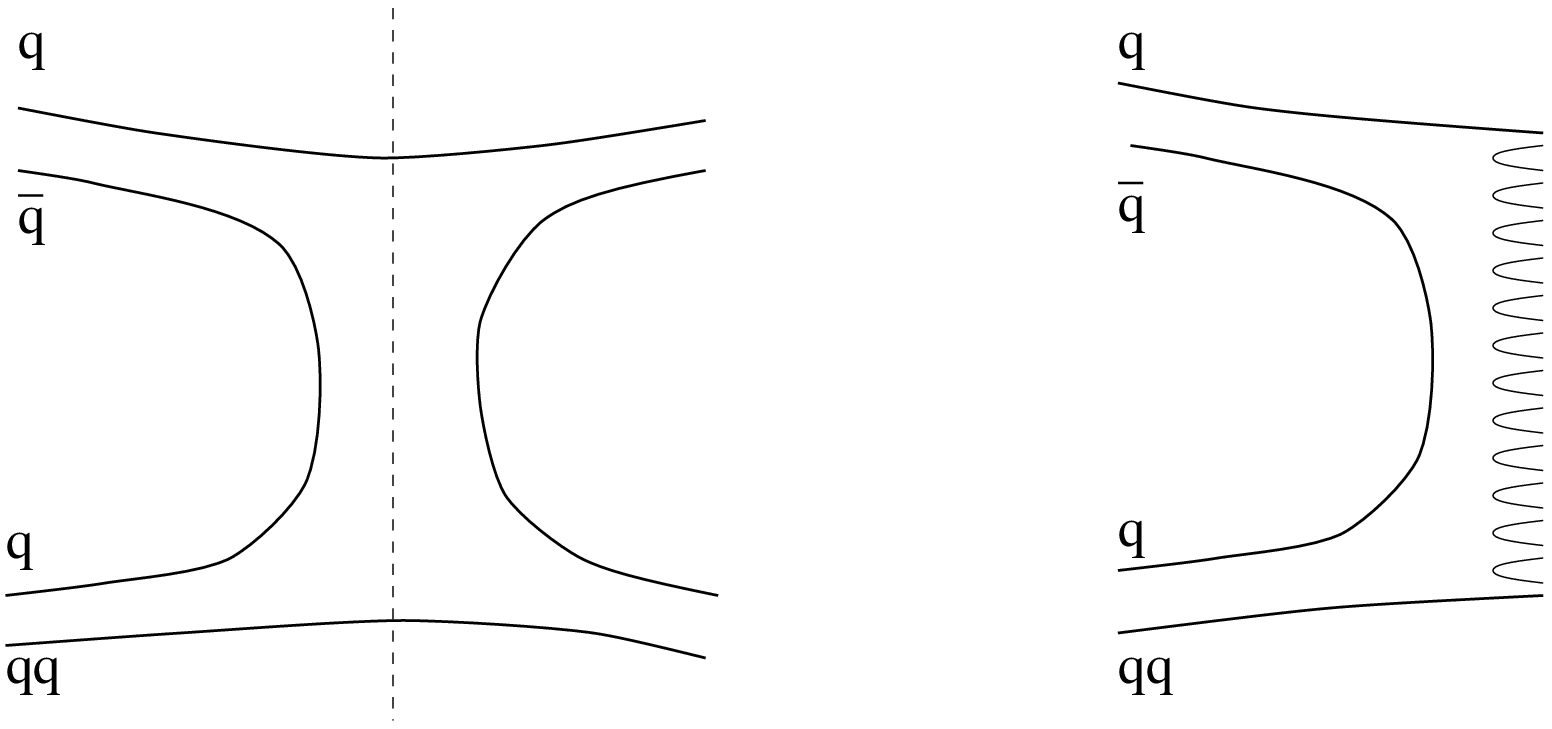,height=50mm,width=110mm}}
\put(3.5,13.7){\bf\large a)}
\put(12.5,13.7){\bf\large b)}
\put(3.0,7.0){\epsfig{figure=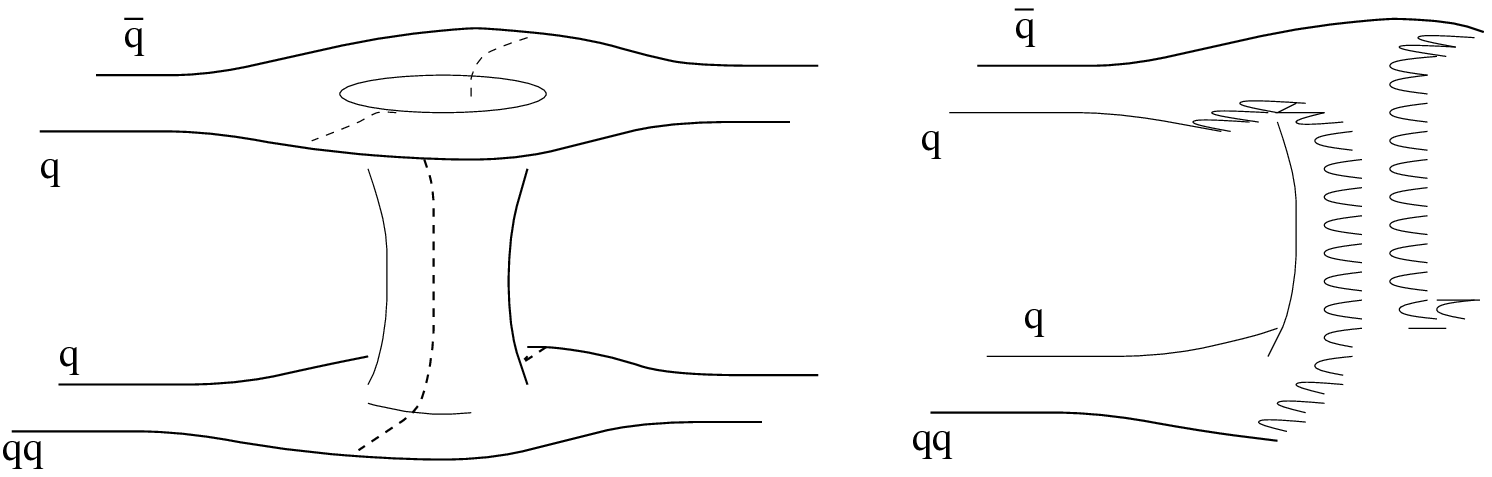,height=55mm,width=140mm}}
\put(3.5,7.0){\bf\large c)}
\put(12.5,7.0){\bf\large d)}
\put(7.5,0.0){\bf\large Fig.~\ref{colflo}}
\end{picture}
\end{figure}
\clearpage
\newpage
\begin{figure}[htb]
\setlength{\unitlength}{1cm}
\begin{picture}(15,23)(0,0)
\put(-2.0,7.5){\psfig{figure=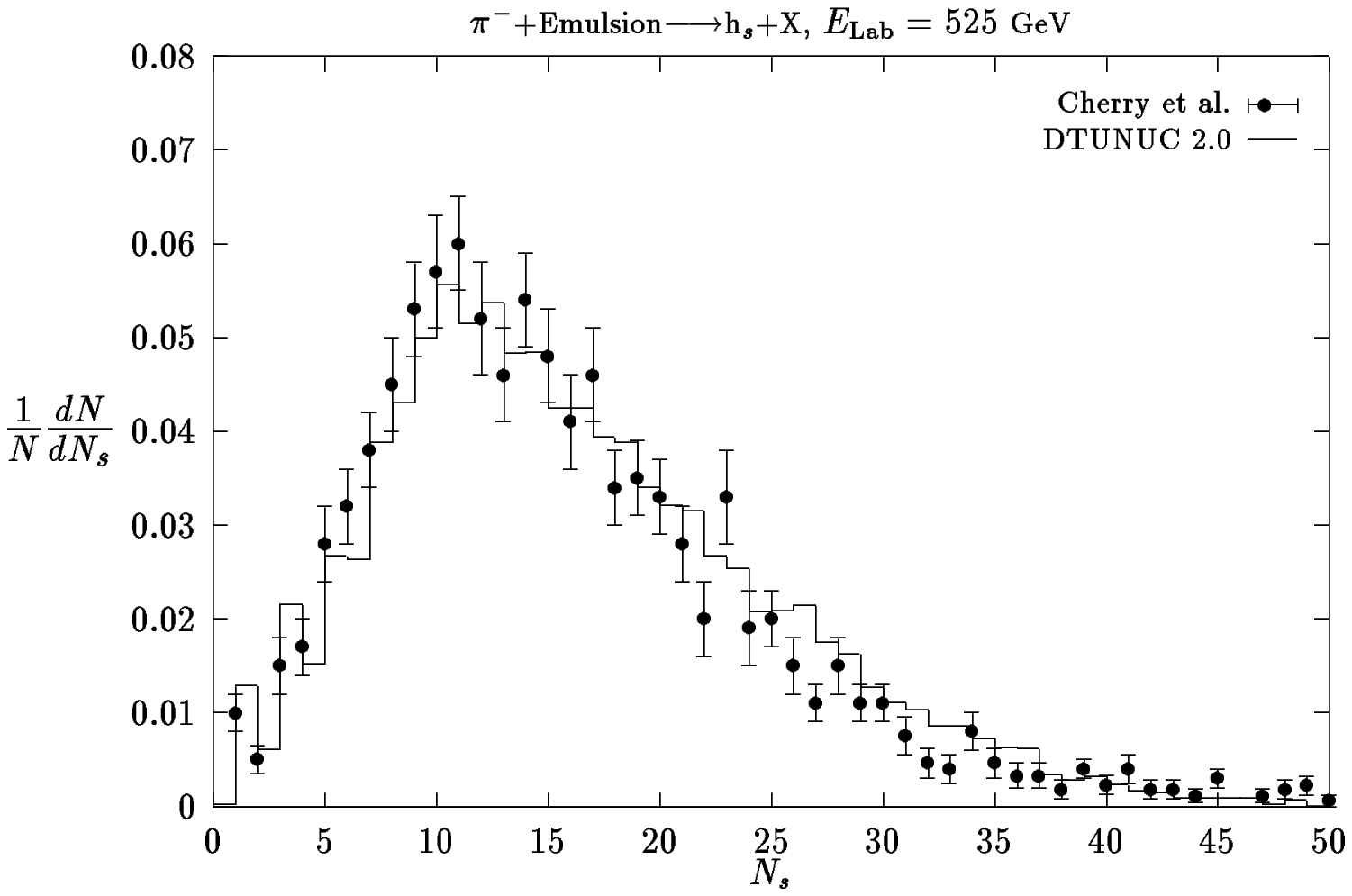}}
\put(1,13.0){\bf\large a)}
\put(-2.0,-4.5){\psfig{figure=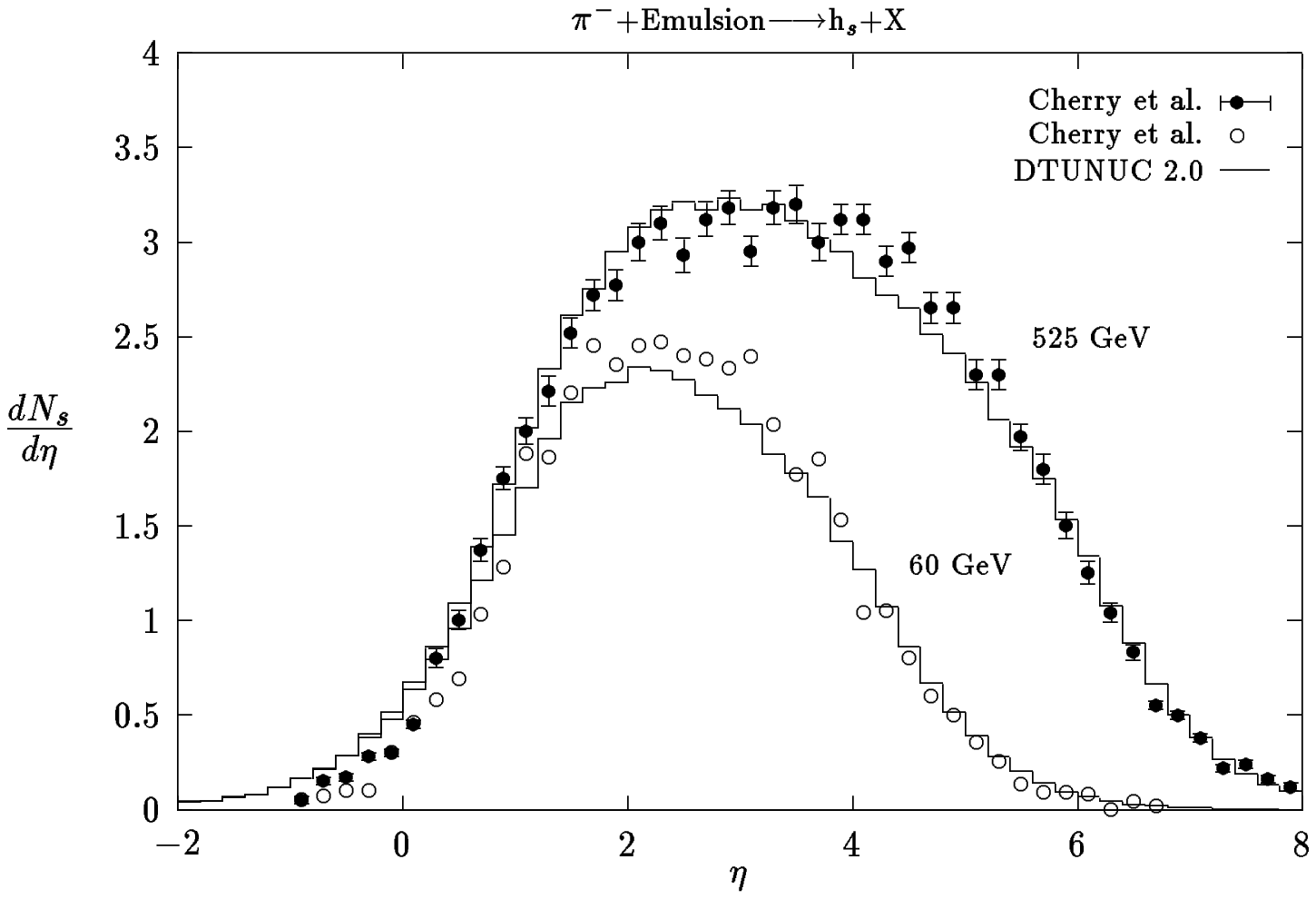}}
\put(1,1.0){\bf\large b)}
\put(7.5,0.0){\bf\large Fig.~\ref{pAmul}}
\end{picture}
\end{figure}
\clearpage
\newpage
\begin{figure}[htb]
\setlength{\unitlength}{1cm}
\begin{picture}(15,23)(0,0)
\put(-2.0,0.0){\psfig{figure=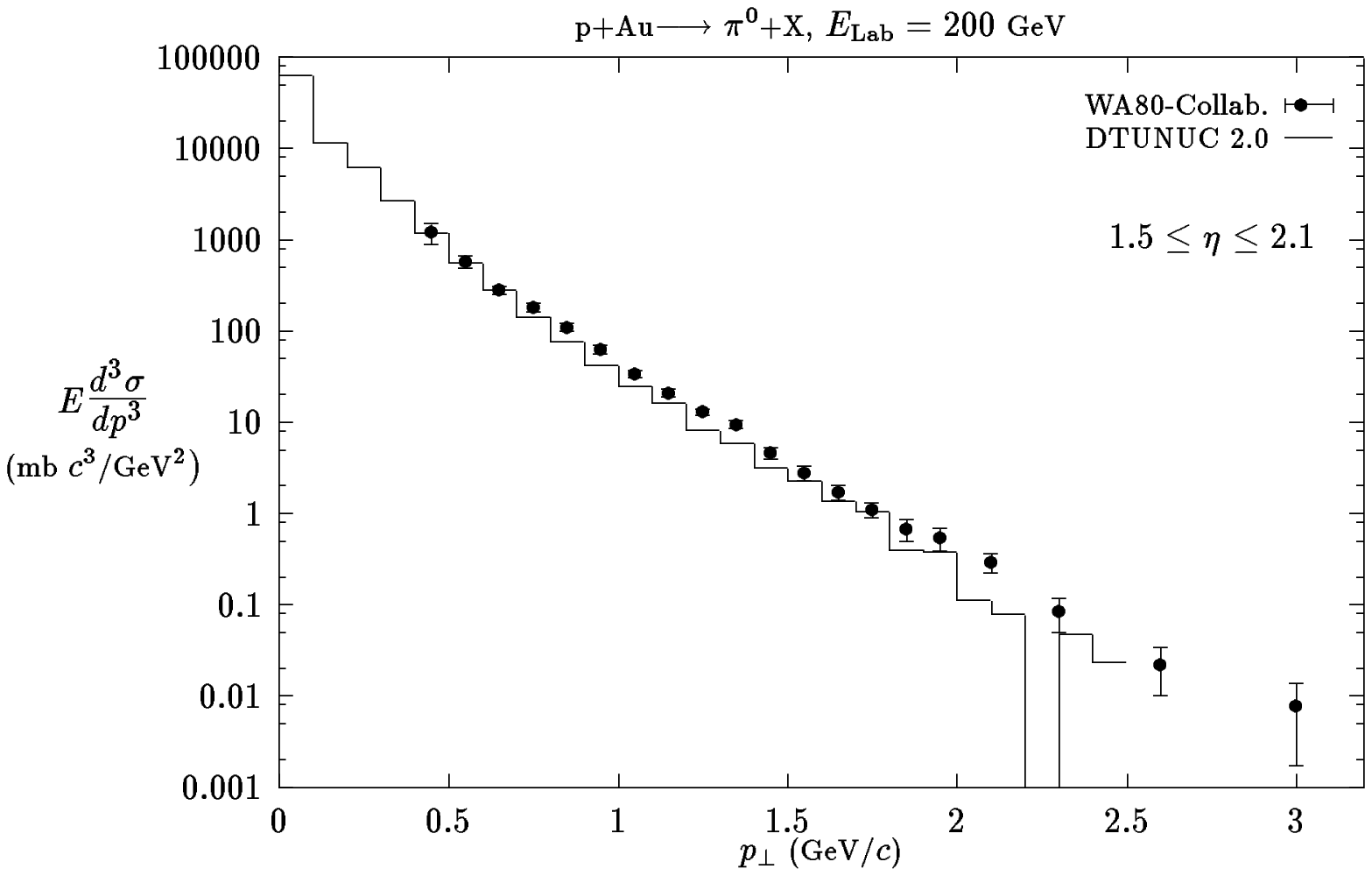}}
\put(7.5,0.0){\bf\large Fig.~\ref{pApt}}
\end{picture}
\end{figure}
\clearpage
\newpage
\begin{figure}[htb]
\setlength{\unitlength}{1cm}
\begin{picture}(15,23)(0,0)
\put(-0.5,10.0){\psfig{figure=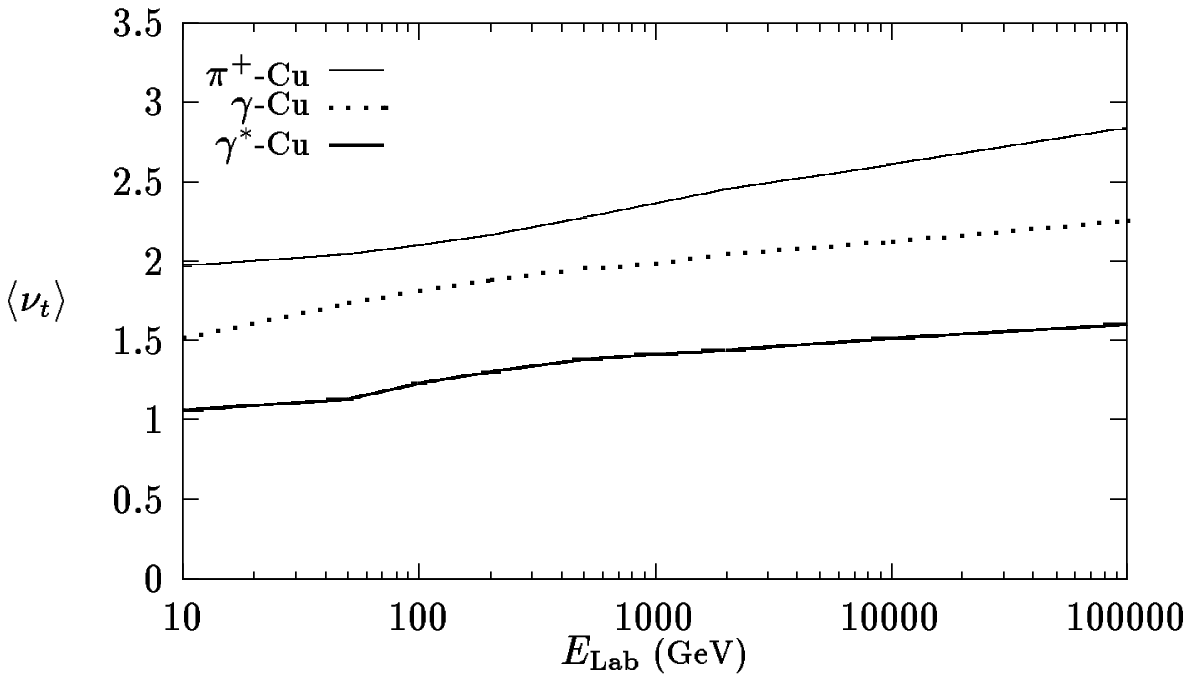}}
\put(3.0,16.0){\bf\large a)}
\put(-0.5,2.5){\psfig{figure=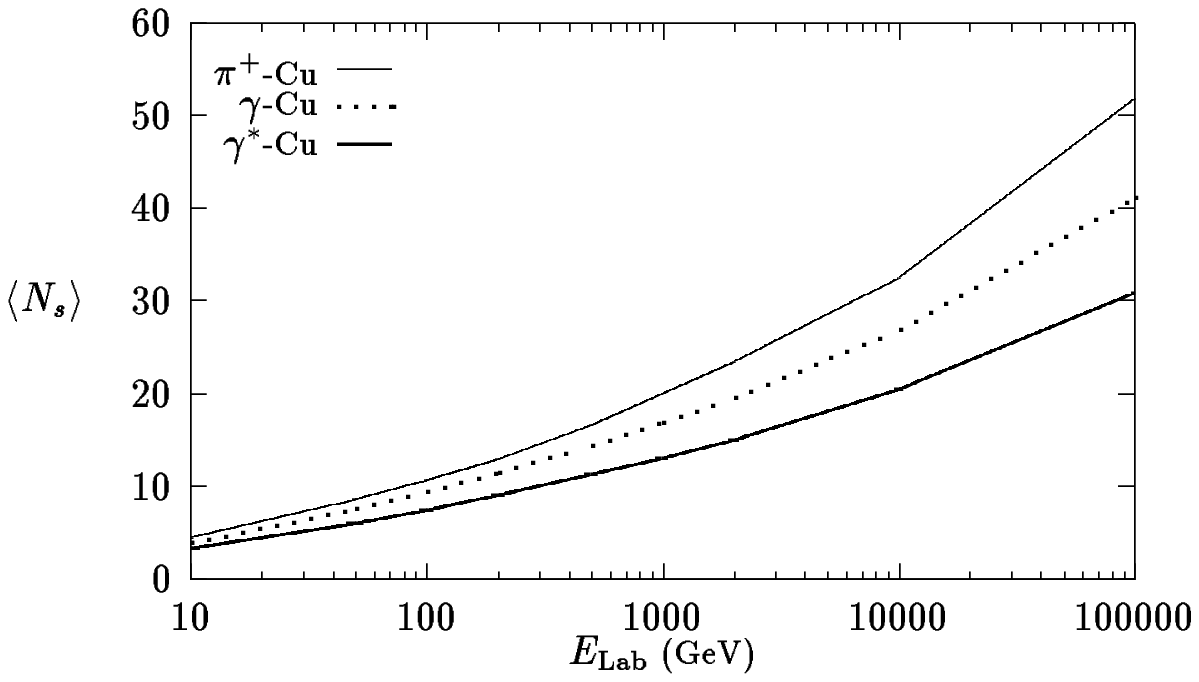}}
\put(3.0,8.5){\bf\large b)}
\put(-0.5,-5.0){\psfig{figure=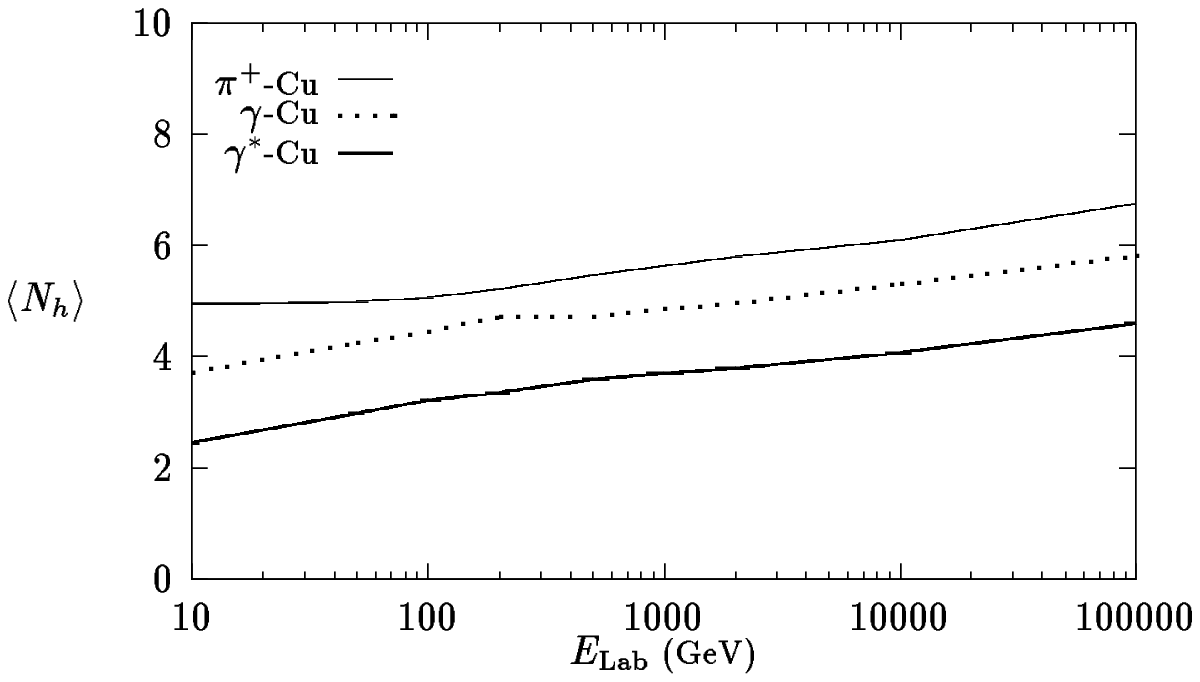}}
\put(3.0,1.0){\bf\large c)}
\put(7.5,0.0){\bf\large Fig.~\ref{iAmult}}
\end{picture}
\end{figure}
\clearpage
\newpage
\begin{figure}[htb]
\setlength{\unitlength}{1cm}
\begin{picture}(15,23)(0,0)
\put(-2.0,7.5){\psfig{figure=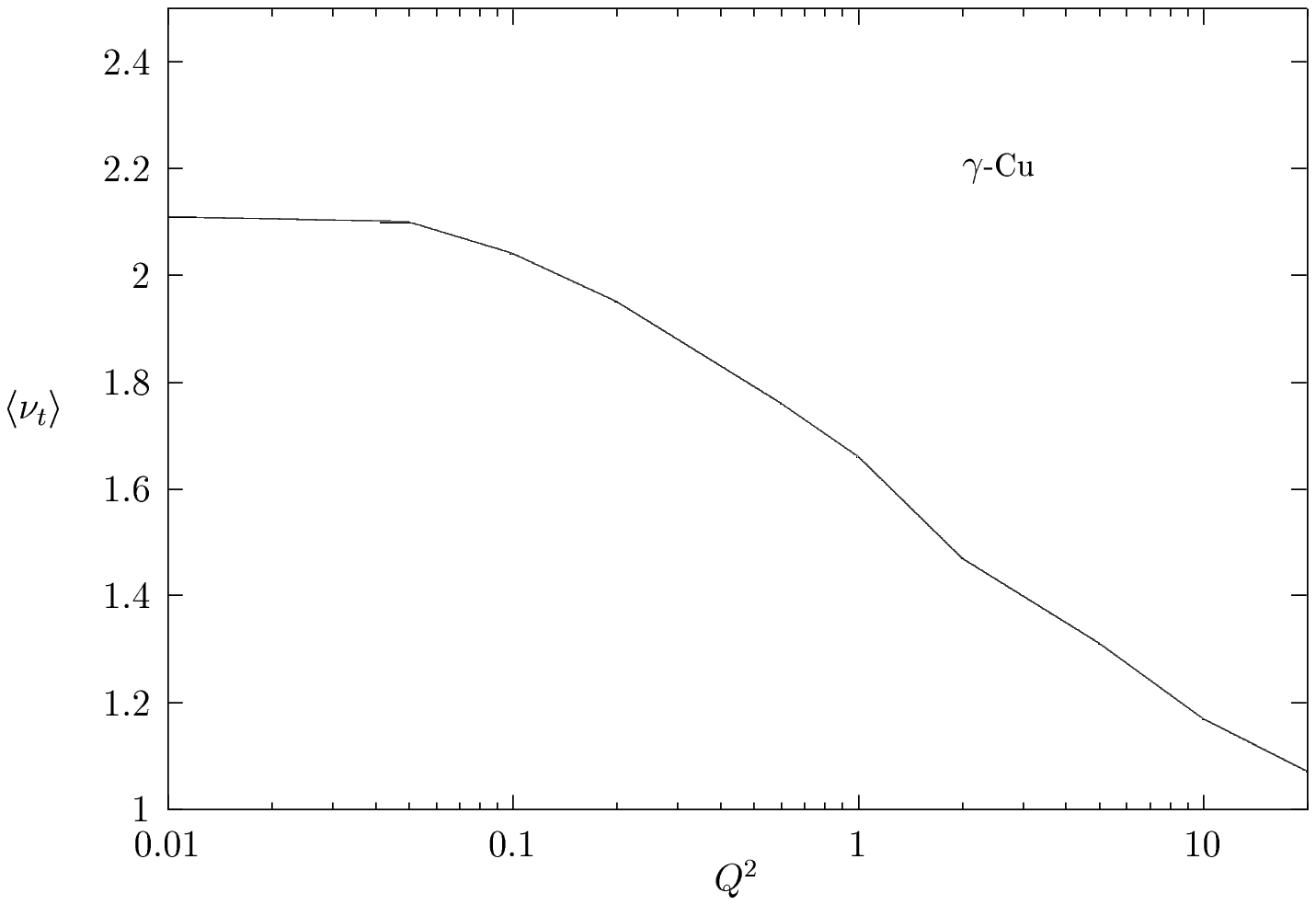}}
\put(1,13.0){\bf\large a)}
\put(-2.0,-4.5){\psfig{figure=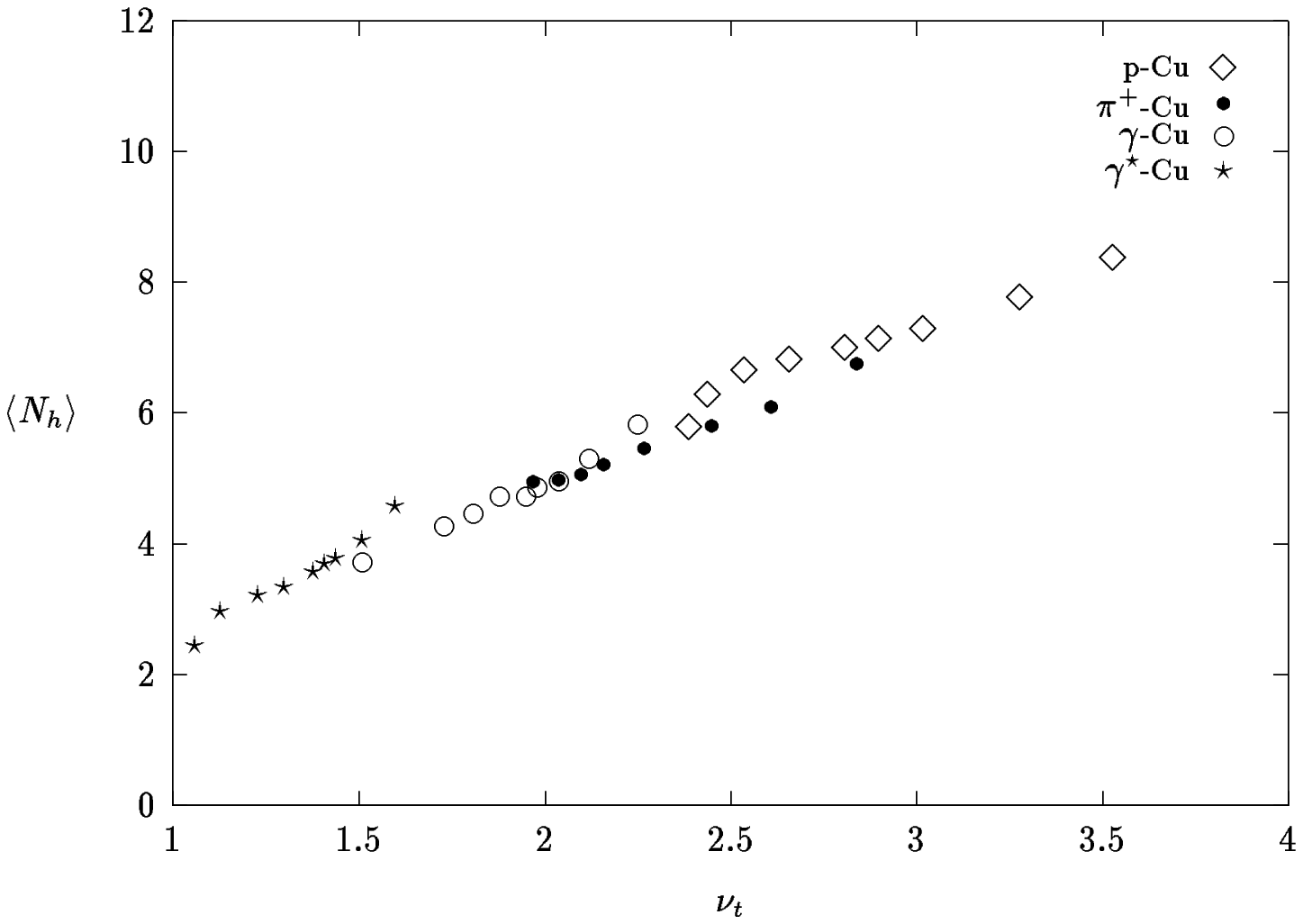}} 
\put(1,1.0){\bf\large b)}
\put(7.5,0.0){\bf\large Fig.~\ref{icunutnh}}
\end{picture}
\end{figure}
\clearpage
\newpage
\begin{figure}[htb]
\setlength{\unitlength}{1cm}
\begin{picture}(15,23)(0,0)
\put(-2.0,0.0){\psfig{figure=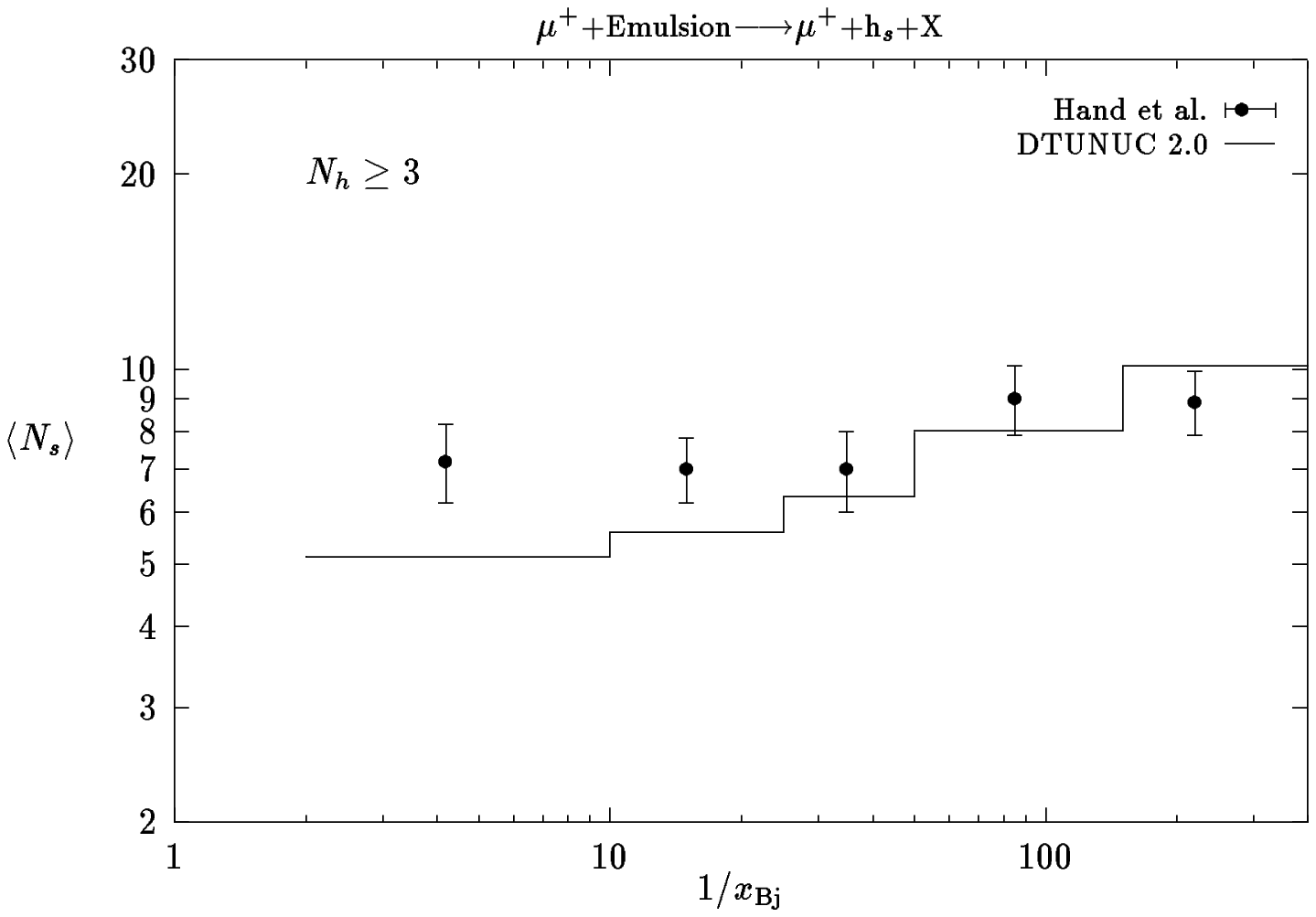}}
\put(7.5,0.0){\bf\large Fig.~\ref{mupemul150nom}}
\end{picture}
\end{figure}
\clearpage
\newpage
\begin{figure}[htb]
\setlength{\unitlength}{1cm}
\begin{picture}(15,23)(0,0)
\put(-4.0,7.5){\psfig{figure=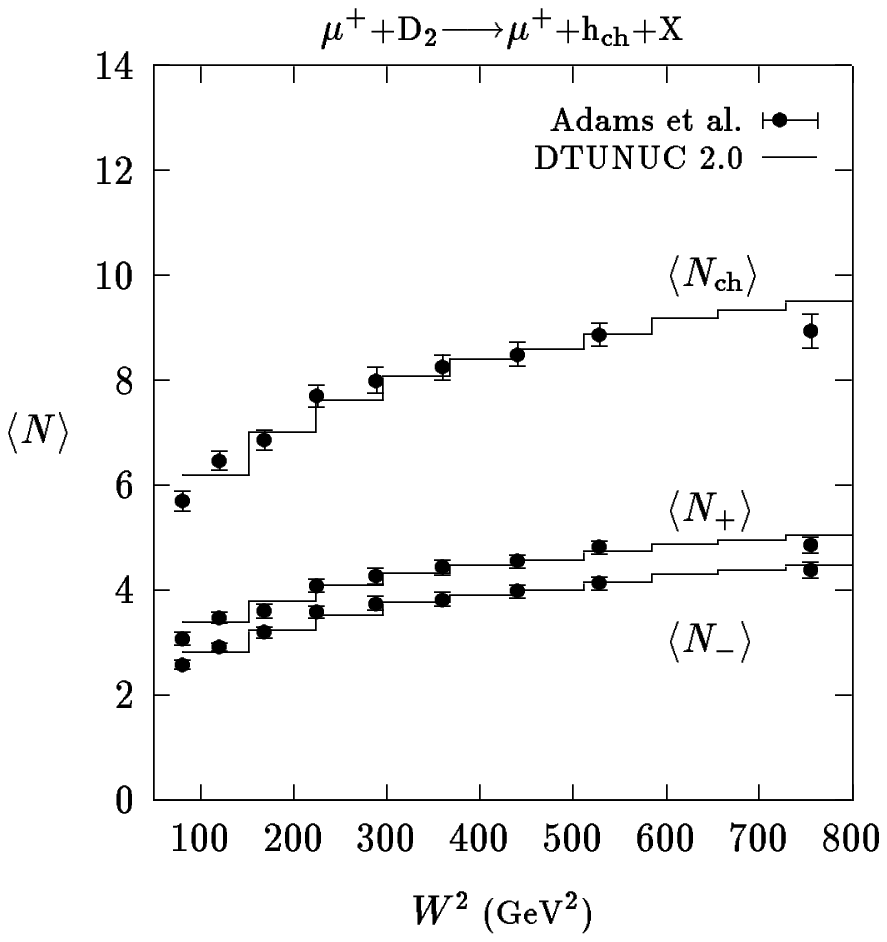}}
\put(1,13.0){\bf\large a)}
\put(5.3,7.5){\psfig{figure=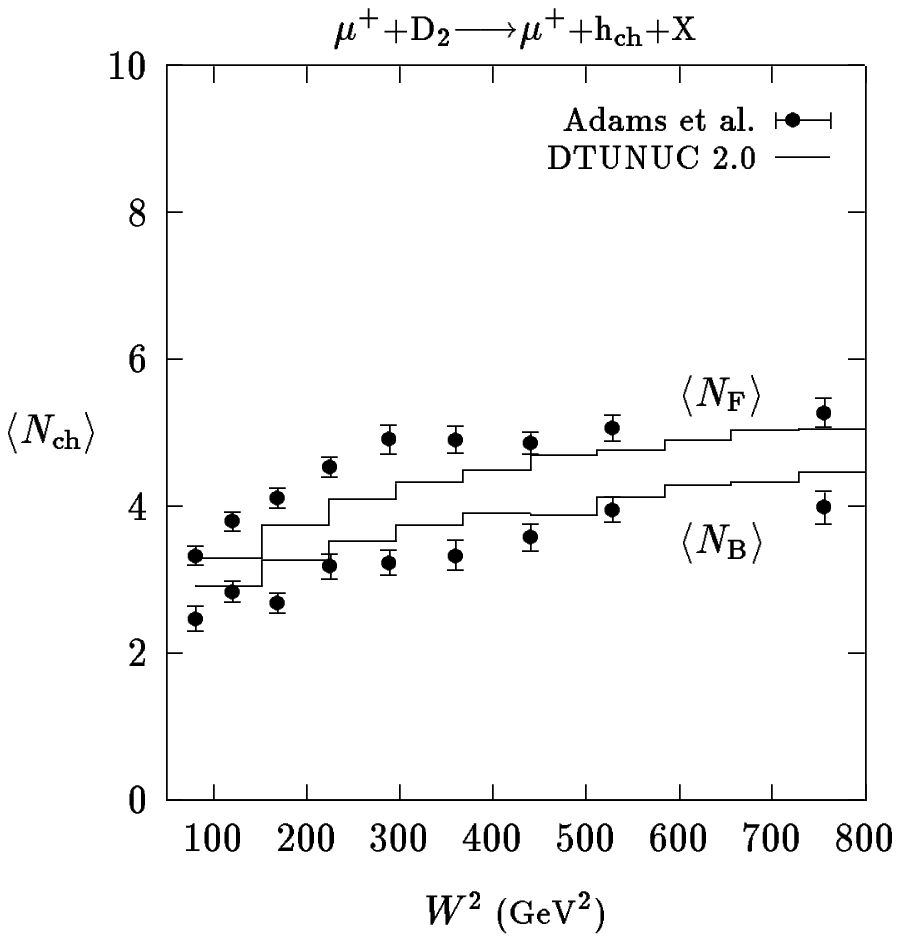}}
\put(10.0,13.0){\bf\large b)}
\put(-4.0,-3.0){\psfig{figure=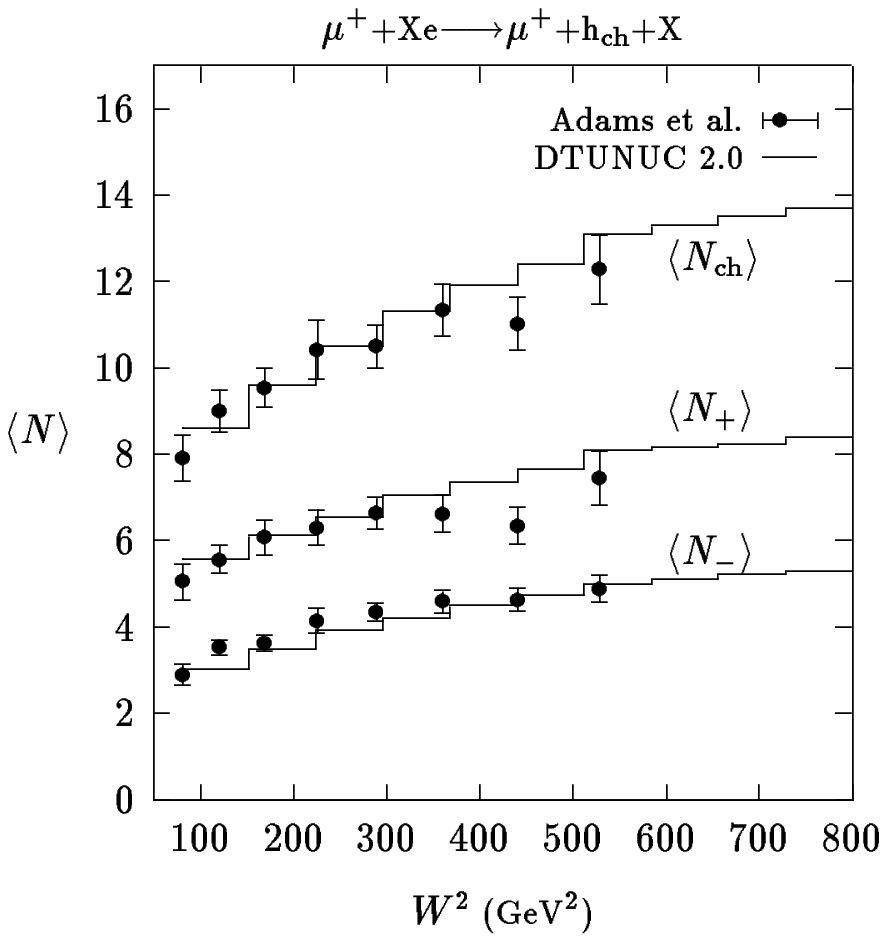}}
\put(1.0,2.5){\bf\large c)}
\put(5.3,-3.0){\psfig{figure=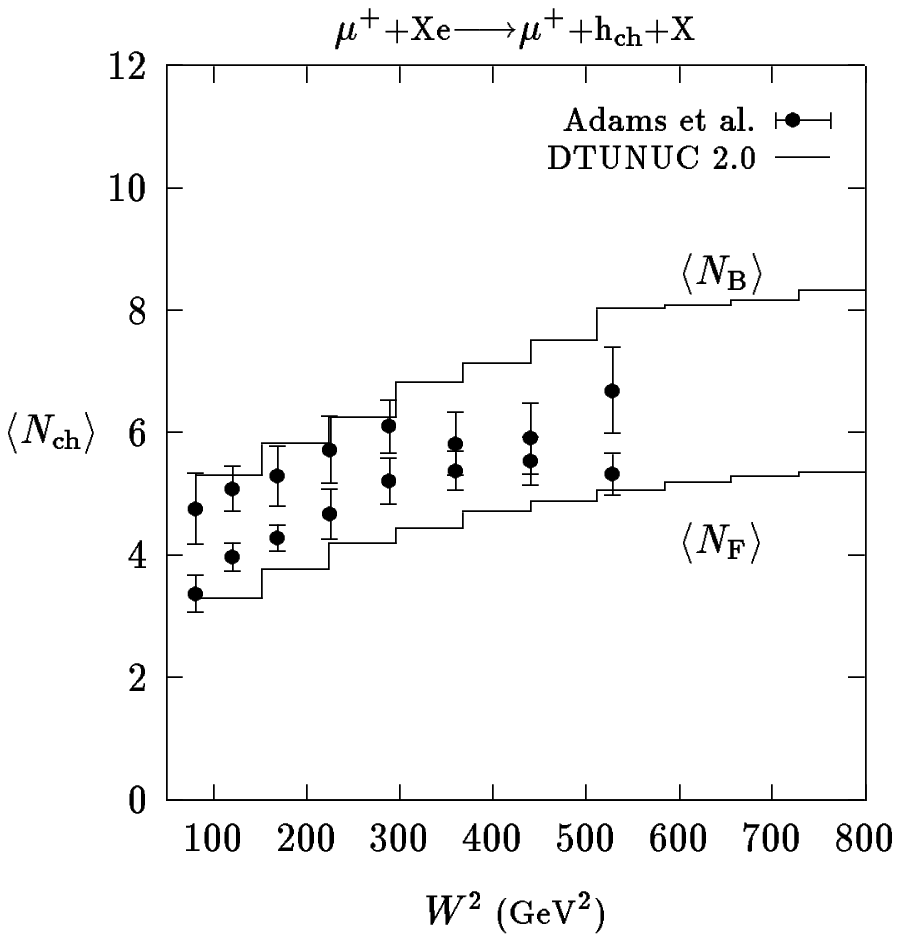}}
\put(10.0,2.5){\bf\large d)}
\put(7.5,0.0){\bf\large Fig.~\ref{mupA490muhpm}}
\end{picture}
\end{figure}
\clearpage
\newpage
\begin{figure}[htb]
\setlength{\unitlength}{1cm}
\begin{picture}(15,23)(0,0)
\put(-2.0,7.5){\psfig{figure=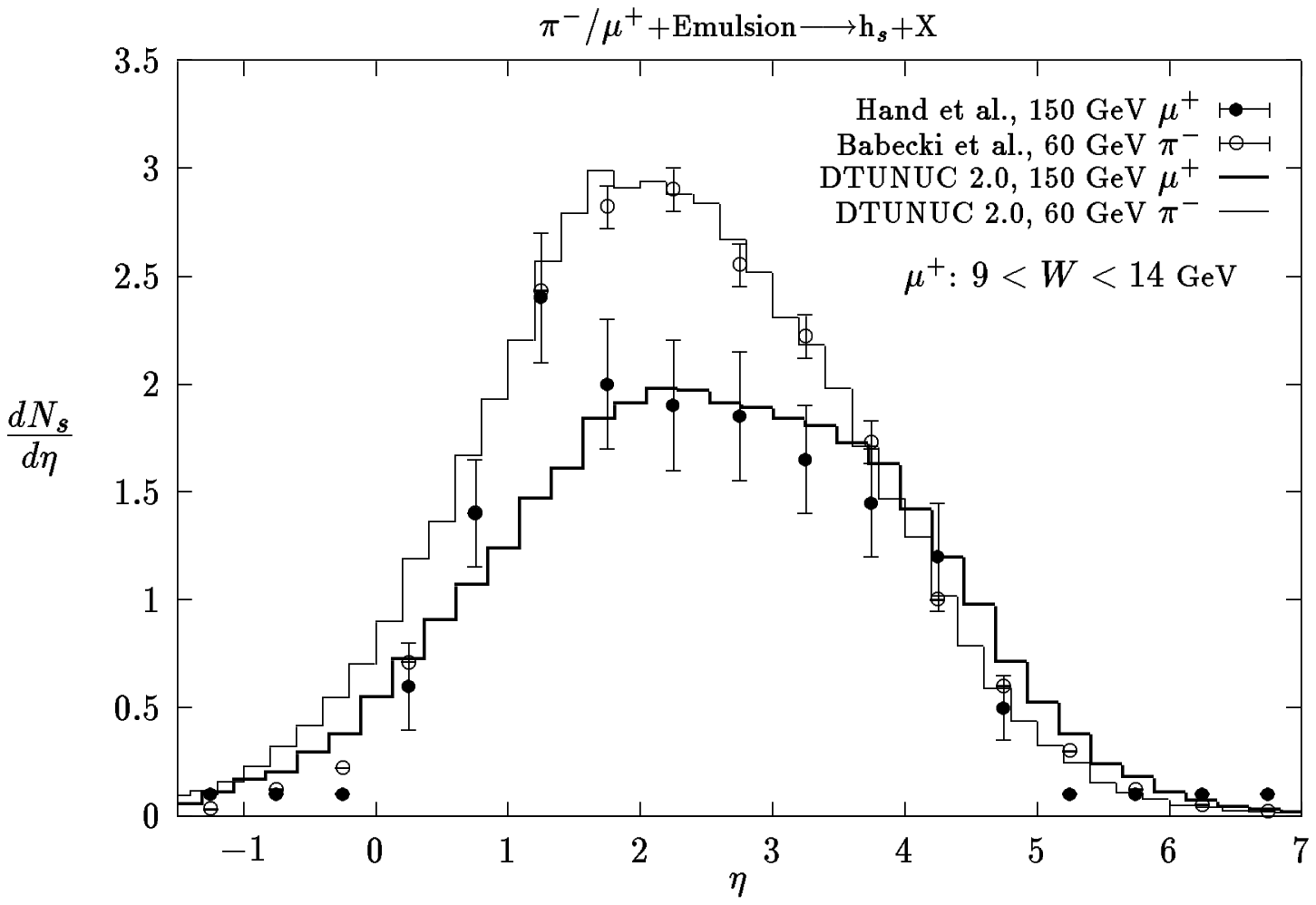}}
\put(1,13.0){\bf\large a)}
\put(-2.0,-4.5){\psfig{figure=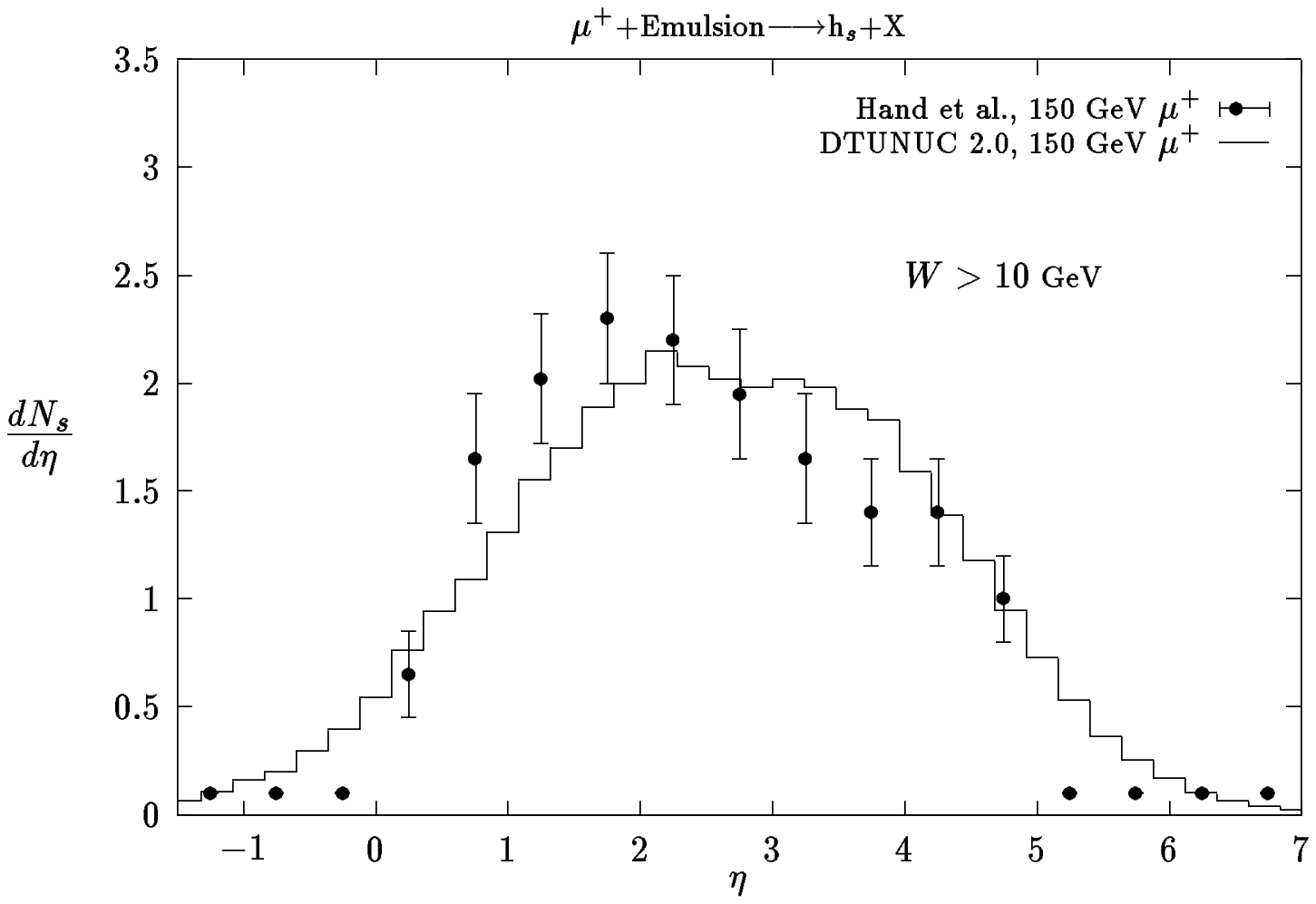}} 
\put(1,1.0){\bf\large b)}
\put(7.5,0.0){\bf\large Fig.~\ref{iemuletans}}
\end{picture}
\end{figure}
\clearpage
\newpage
\begin{figure}[htb]
\setlength{\unitlength}{1cm}
\begin{picture}(15,23)(0,0)
\put(-1.0,16.22){\input{mupdeu490yyhp1.pic}}
\put(-1.0, 9.0){\input{mupdeu490yyhp2.pic}}
\put(-1.0,1.78){\input{mupdeu490yyhp3.pic}}
\put(4.3,1.0){\bf\large a)}
\put(8.5,16.22){\input{mupdeu490yyhm1.pic}}
\put(8.5, 9.0){\input{mupdeu490yyhm2.pic}}
\put(8.5,1.78){\input{mupdeu490yyhm3.pic}}
\put(13.8,1.0){\bf\large b)}
\put(7.5,0.0){\bf\large Fig.~\ref{mupdeu490yyhpm}}
\end{picture}
\end{figure}
\clearpage
\newpage
\begin{figure}[htb]
\setlength{\unitlength}{1cm}
\begin{picture}(15,23)(0,0)
\put(-1.0,16.22){\input{mupxe490yyhp1.pic}}
\put(-1.0, 9.0){\input{mupxe490yyhp2.pic}}
\put(-1.0,1.78){\input{mupxe490yyhp3.pic}}
\put(4.3,1.0){\bf\large a)}
\put(8.5,16.22){\input{mupxe490yyhm1.pic}}
\put(8.5, 9.0){\input{mupxe490yyhm2.pic}}
\put(8.5,1.78){\input{mupxe490yyhm3.pic}}
\put(13.8,1.0){\bf\large b)}
\put(7.5,0.0){\bf\large Fig.~\ref{mupxe490yyhpm}}
\end{picture}
\end{figure}
\clearpage
\newpage
\begin{figure}[htb]
\setlength{\unitlength}{1cm}
\begin{picture}(15,23)(0,0)
\put(-3.5,7.5){\psfig{figure=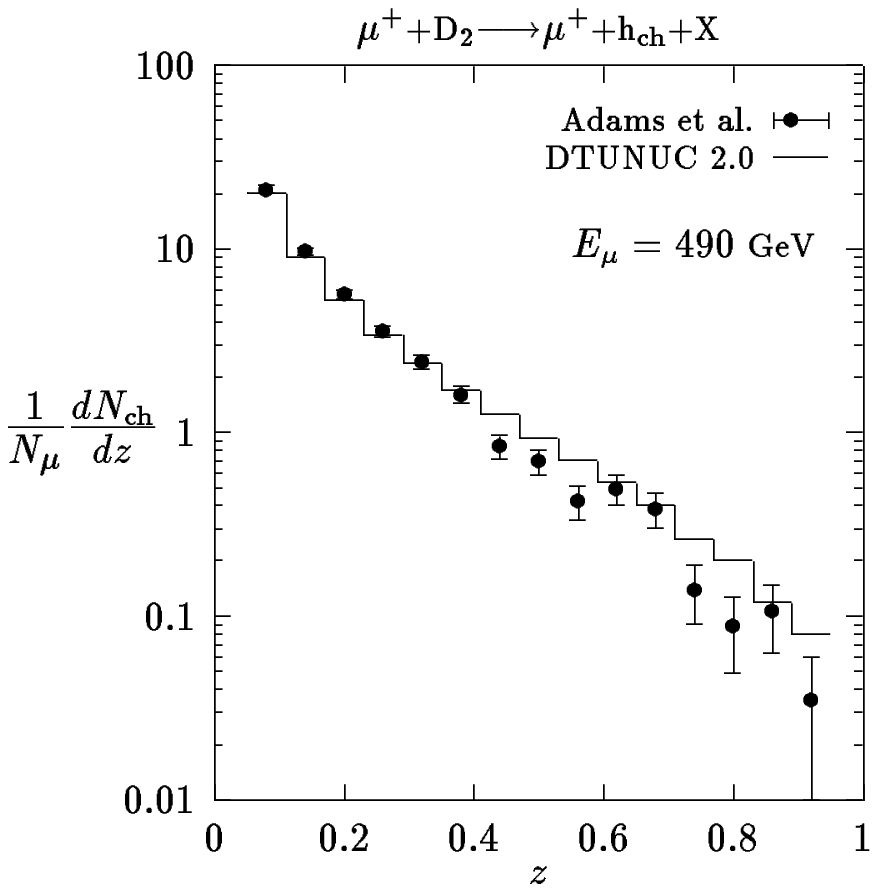}}
\put(1.0,13.0){\bf\large a)}
\put(6.0,7.5){\psfig{figure=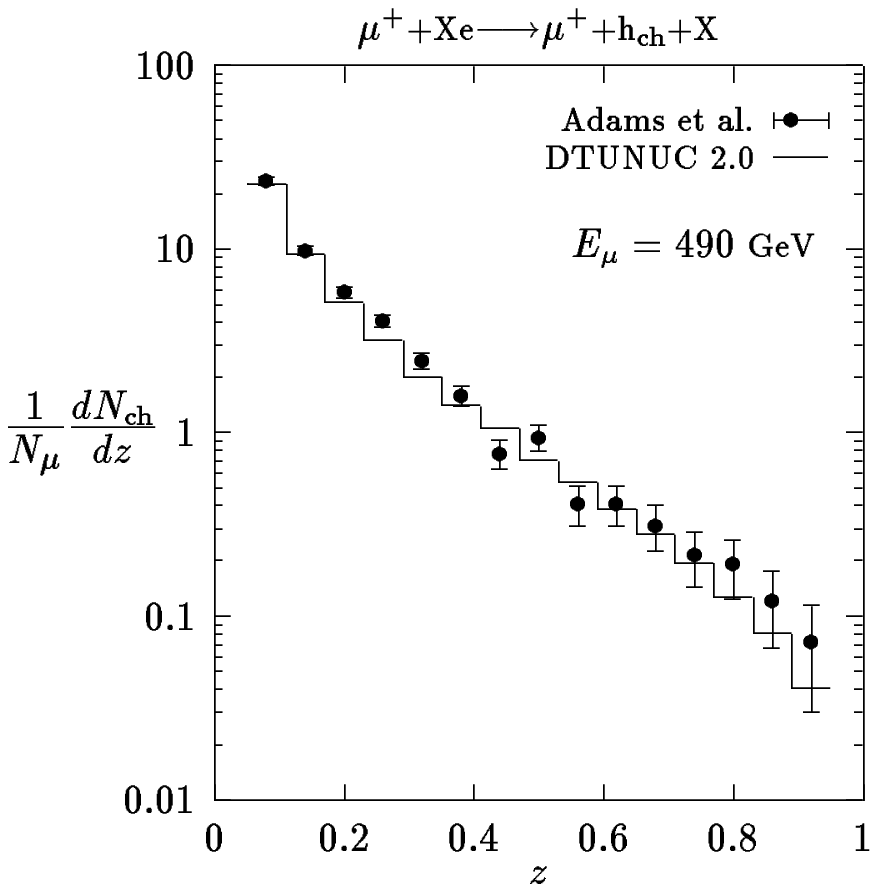}}
\put(10.0,13.0){\bf\large b)}
\put(-2.0,-4.0){\psfig{figure=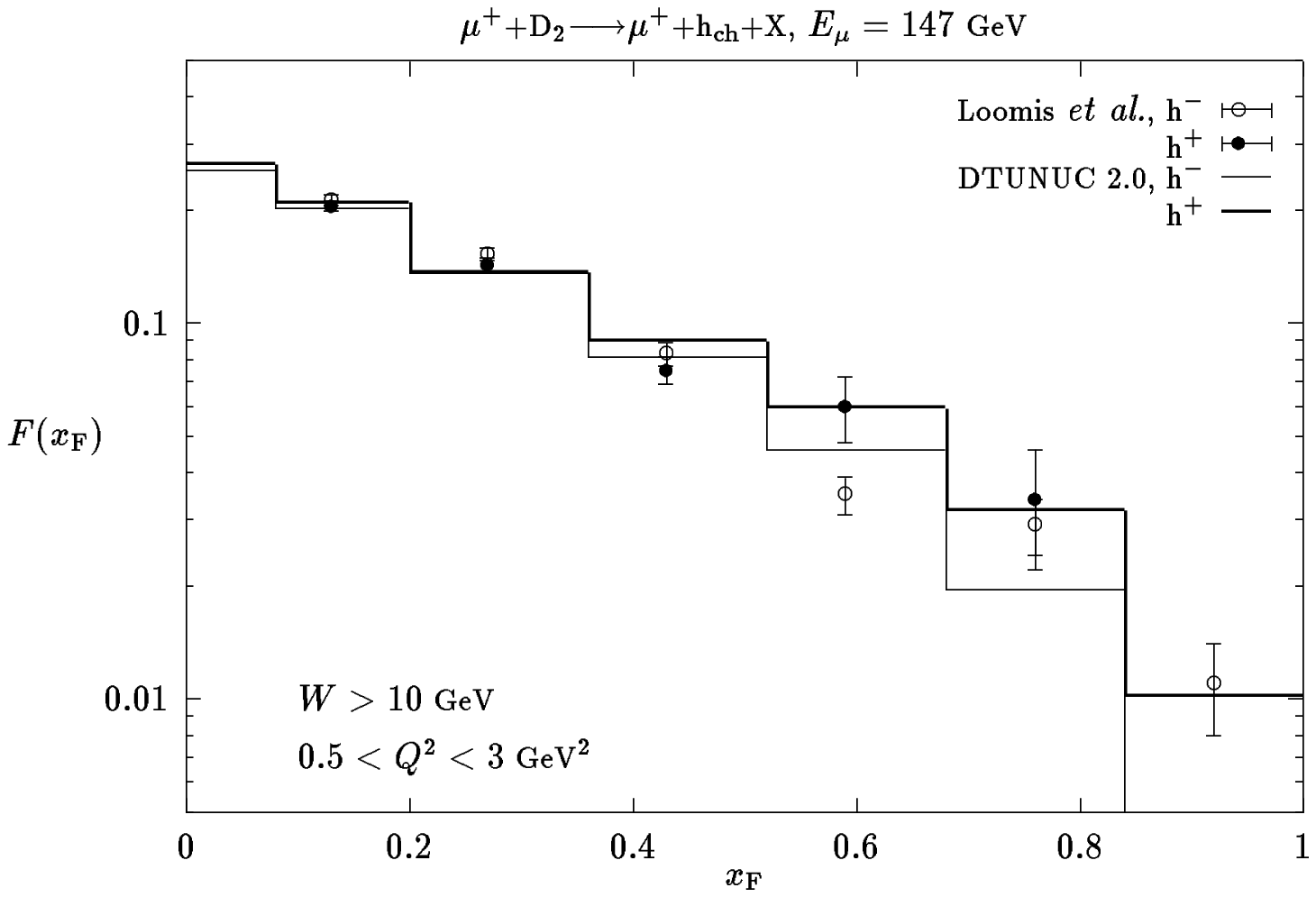}}
\put(1.0,1.5){\bf\large c)}
\put(7.5,0.0){\bf\large Fig.~\ref{mupA490zhpm}}
\end{picture}
\end{figure}
\clearpage
\newpage
\begin{figure}[htb]
\setlength{\unitlength}{1cm}
\begin{picture}(15,23)(0,0)
\put(-2.0,7.5){\psfig{figure=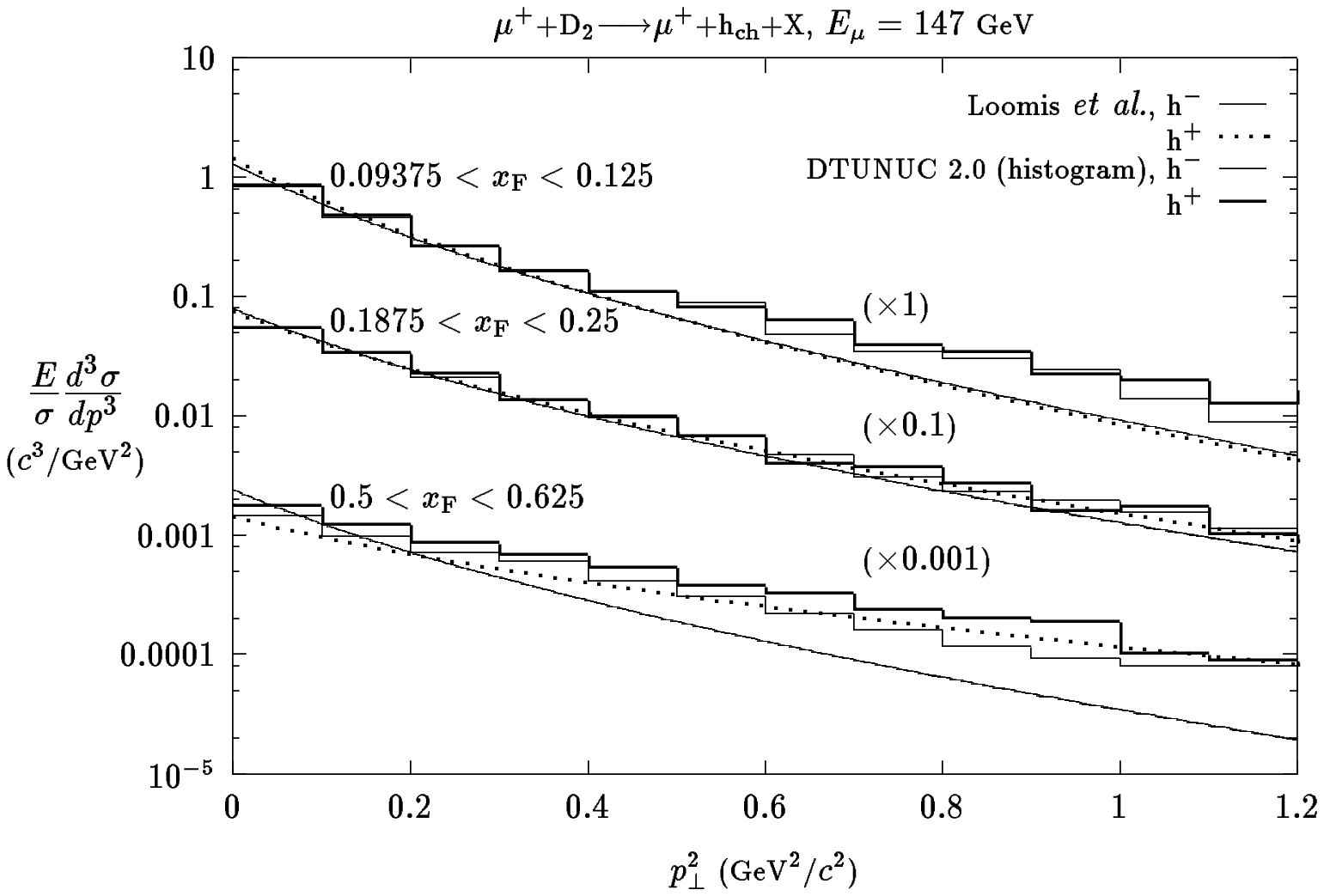}}
\put(1,13.0){\bf\large a)}
\put(-2.0,-4.5){\psfig{figure=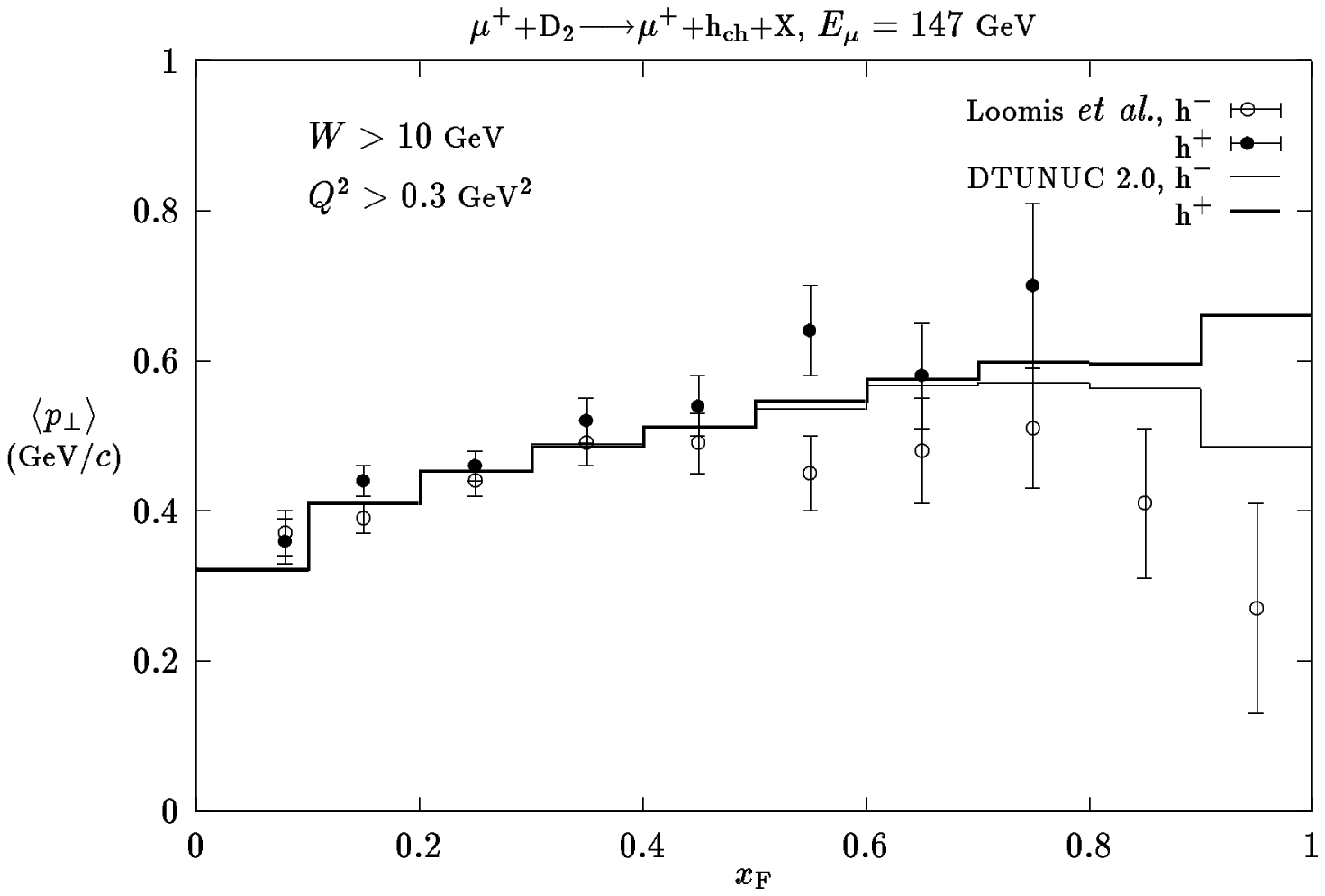}}
\put(1,1.0){\bf\large b)}
\put(7.5,0.0){\bf\large Fig.~\ref{mupdeu147pt}}
\end{picture}
\end{figure}
\clearpage
\newpage
\begin{figure}[htb]
\setlength{\unitlength}{1cm}
\begin{picture}(15,23)(0,0)
\put(-0.5,10.0){\psfig{figure=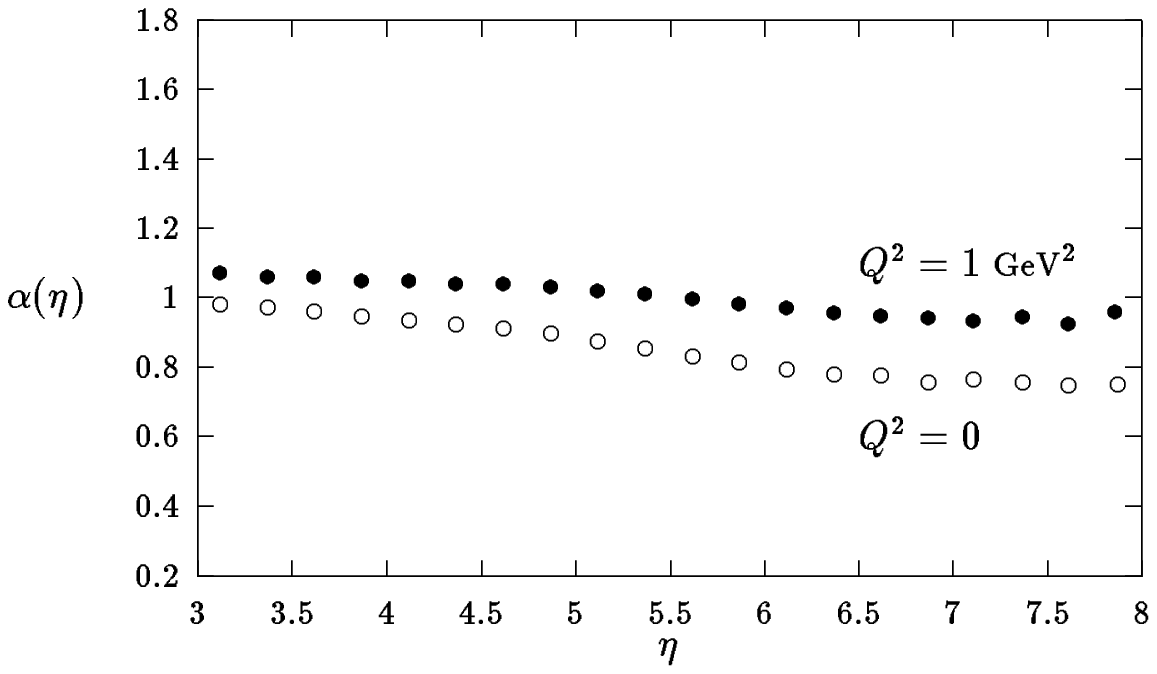}}
\put(3.0,16.0){\bf\large a)}
\put(-0.5,2.5){\psfig{figure=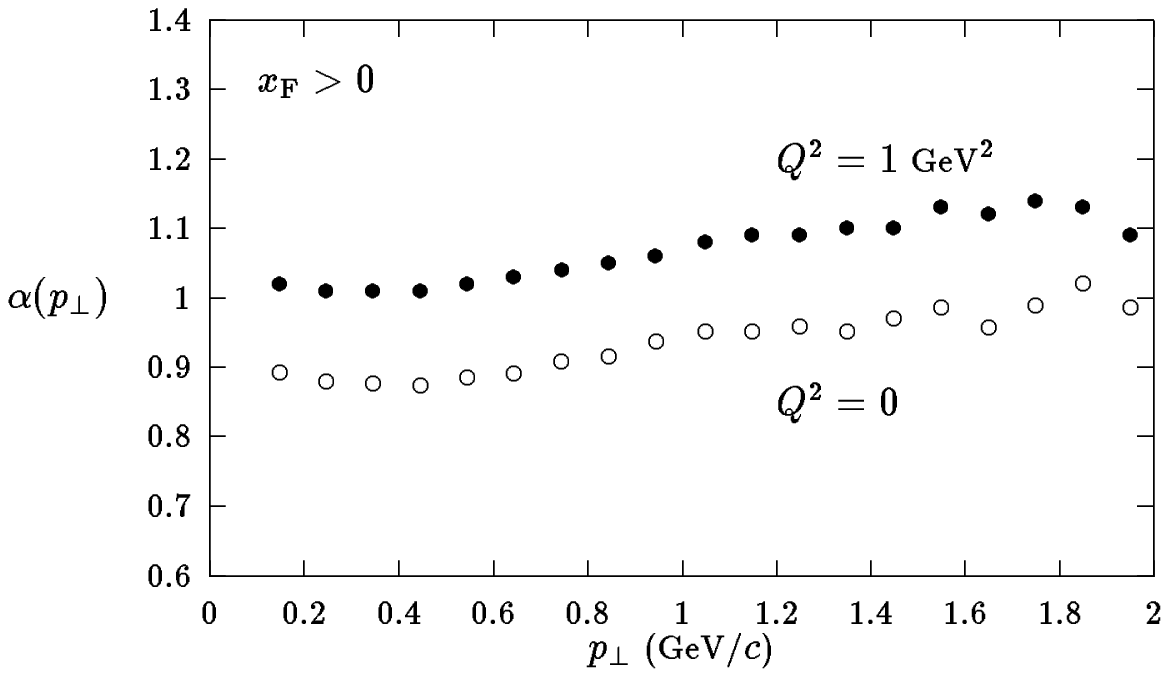}}
\put(3.0,8.5){\bf\large b)}
\put(-0.5,-5.0){\psfig{figure=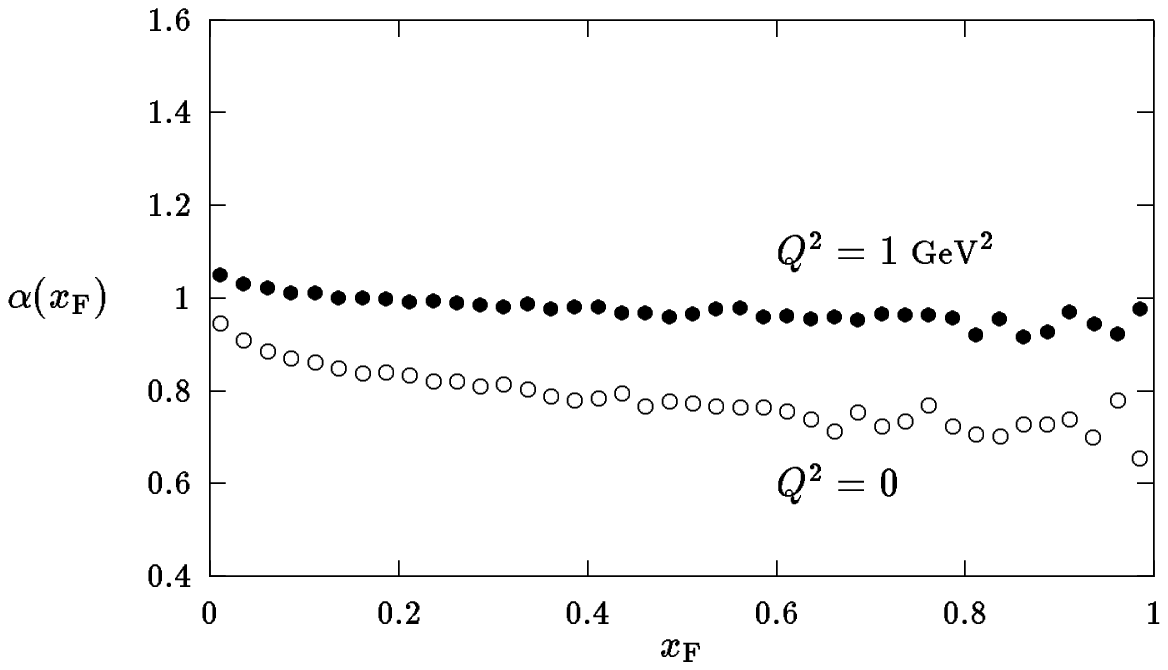}}
\put(3.0,1.0){\bf\large c)}
\put(7.5,0.0){\bf\large Fig.~\ref{iAalp}}
\end{picture}
\end{figure}
\clearpage
\newpage
\begin{figure}[htb]
\setlength{\unitlength}{1cm}
\begin{picture}(15,23)(0,0)
\put(-2.0,0.0){\psfig{figure=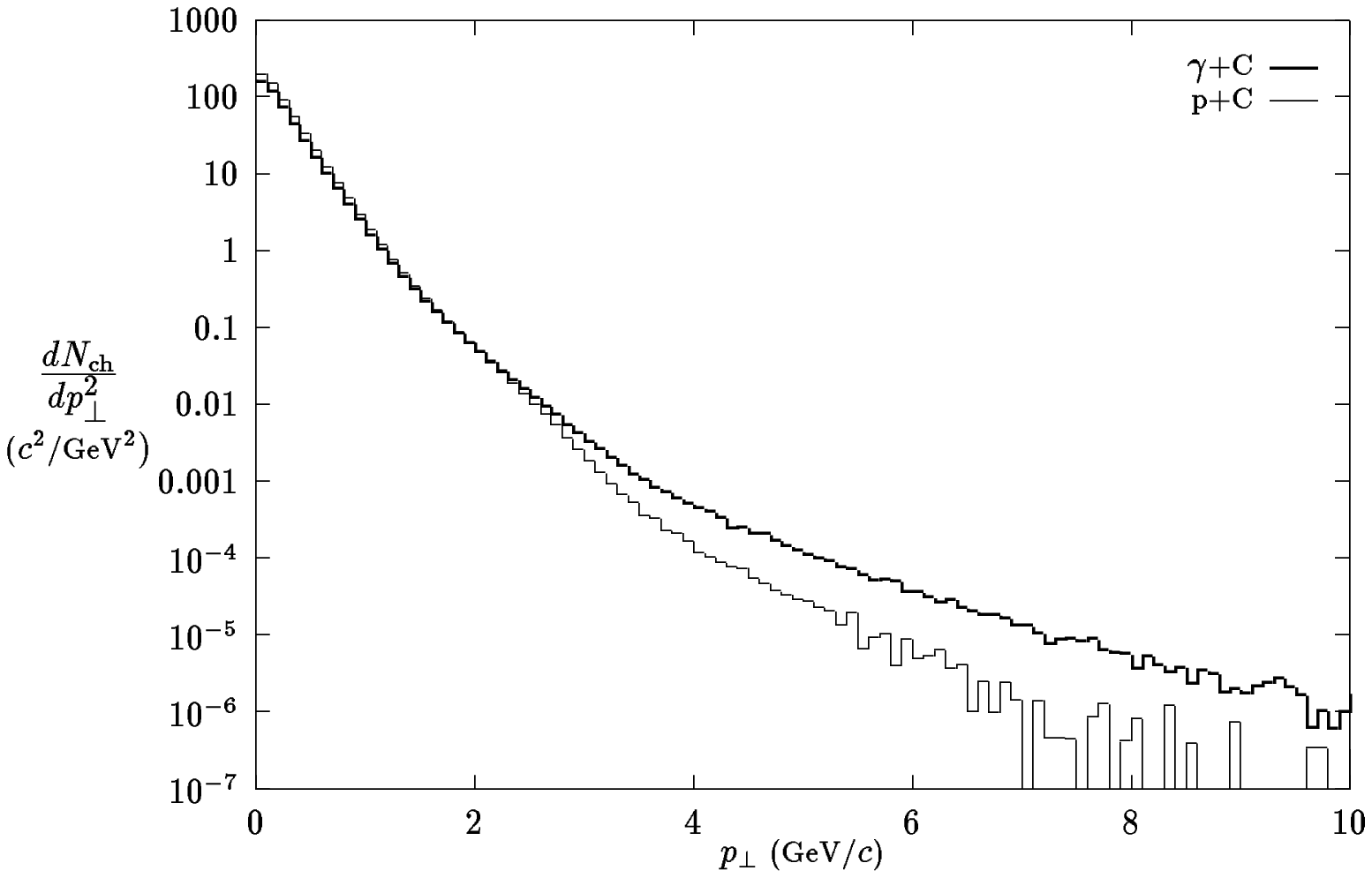}}
\put(7.5,0.0){\bf\large Fig.~\ref{iiherapt}}
\end{picture}
\end{figure}
\clearpage
\newpage
\begin{figure}[htb]
\setlength{\unitlength}{1cm}
\begin{picture}(15,23)(0,0)
\put(-2.0,7.5){\psfig{figure=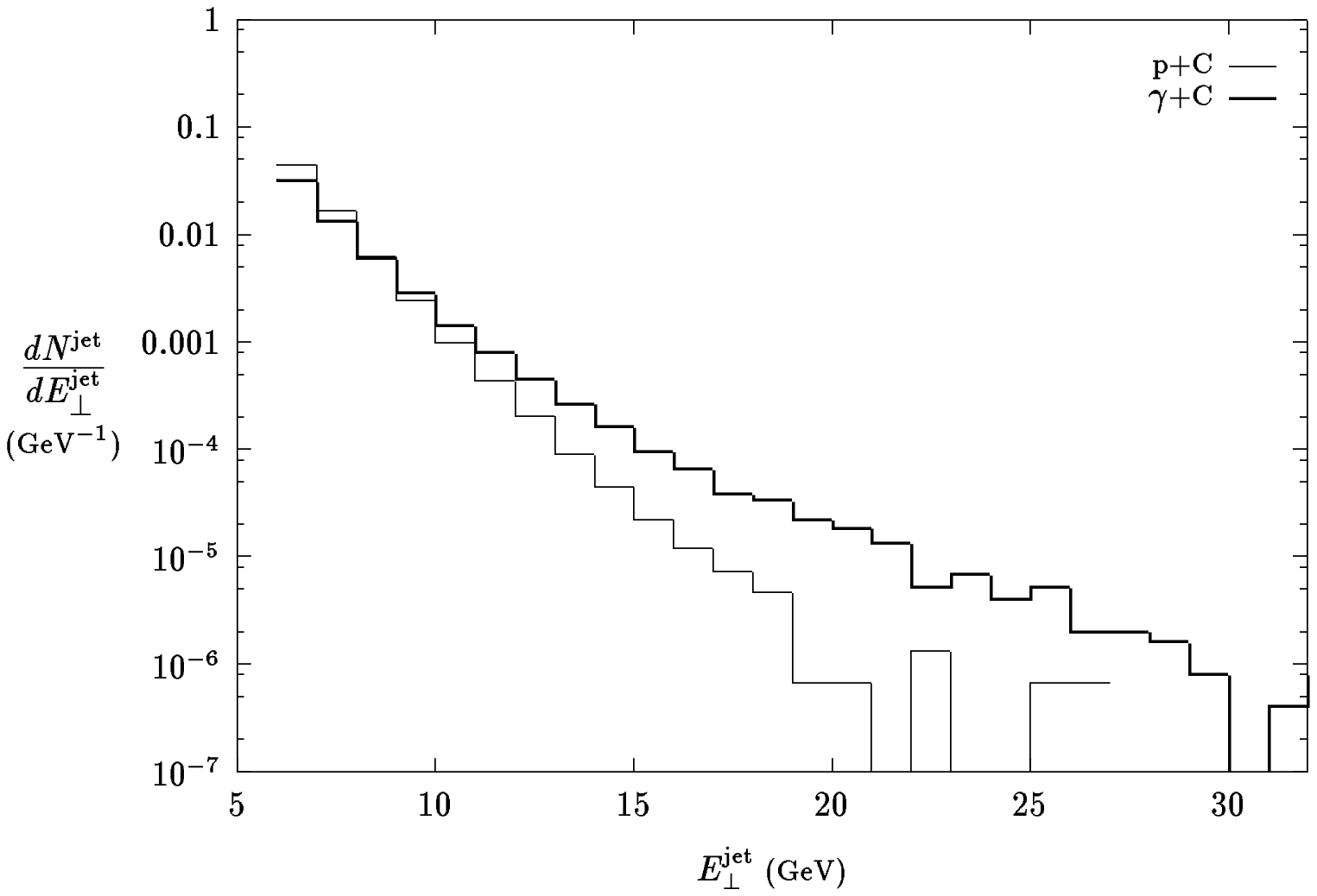}}
\put(1,13.0){\bf\large a)}
\put(-2.0,-4.5){\psfig{figure=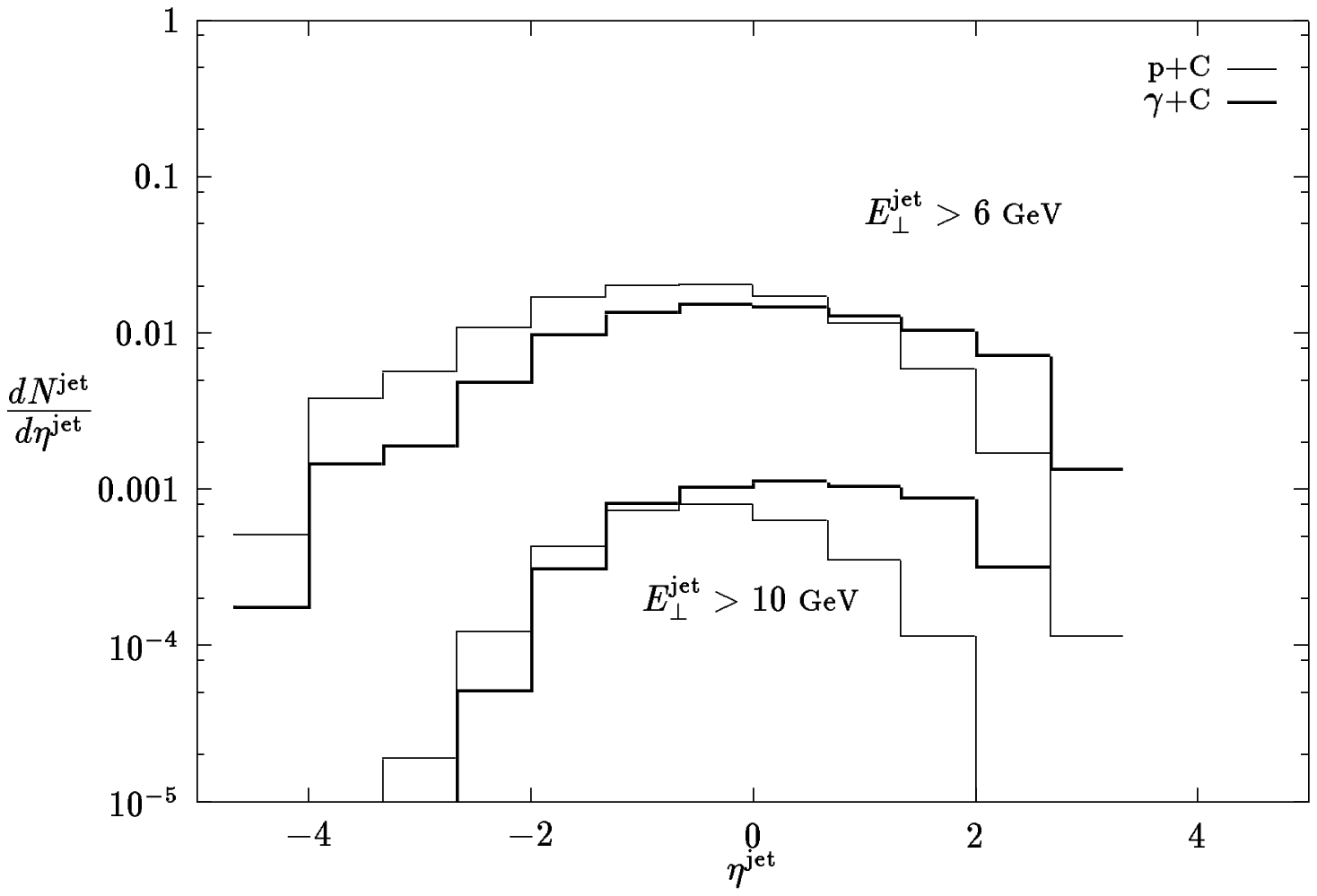}}
\put(1,1.0){\bf\large b)}
\put(7.5,0.0){\bf\large Fig.~\ref{iAjet}}
\end{picture}
\end{figure}
\clearpage
\newpage
\begin{figure}[htb]
\setlength{\unitlength}{1cm}
\begin{picture}(15,23)(0,0)
\put(-2.0,7.8){\psfig{figure=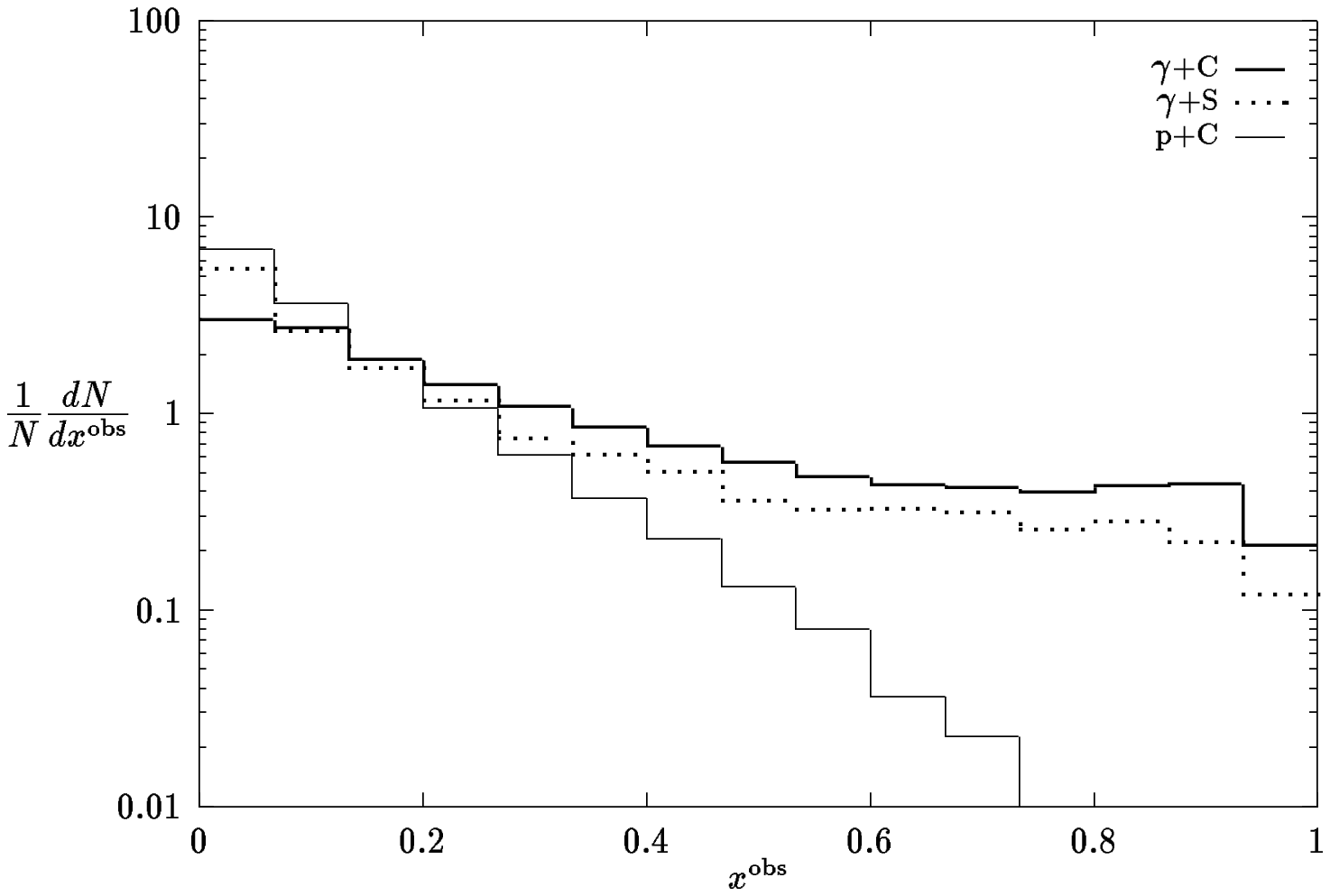}}
\put(1,14.0){\bf\large a)}
\put(-4.0,-3.9){\psfig{figure=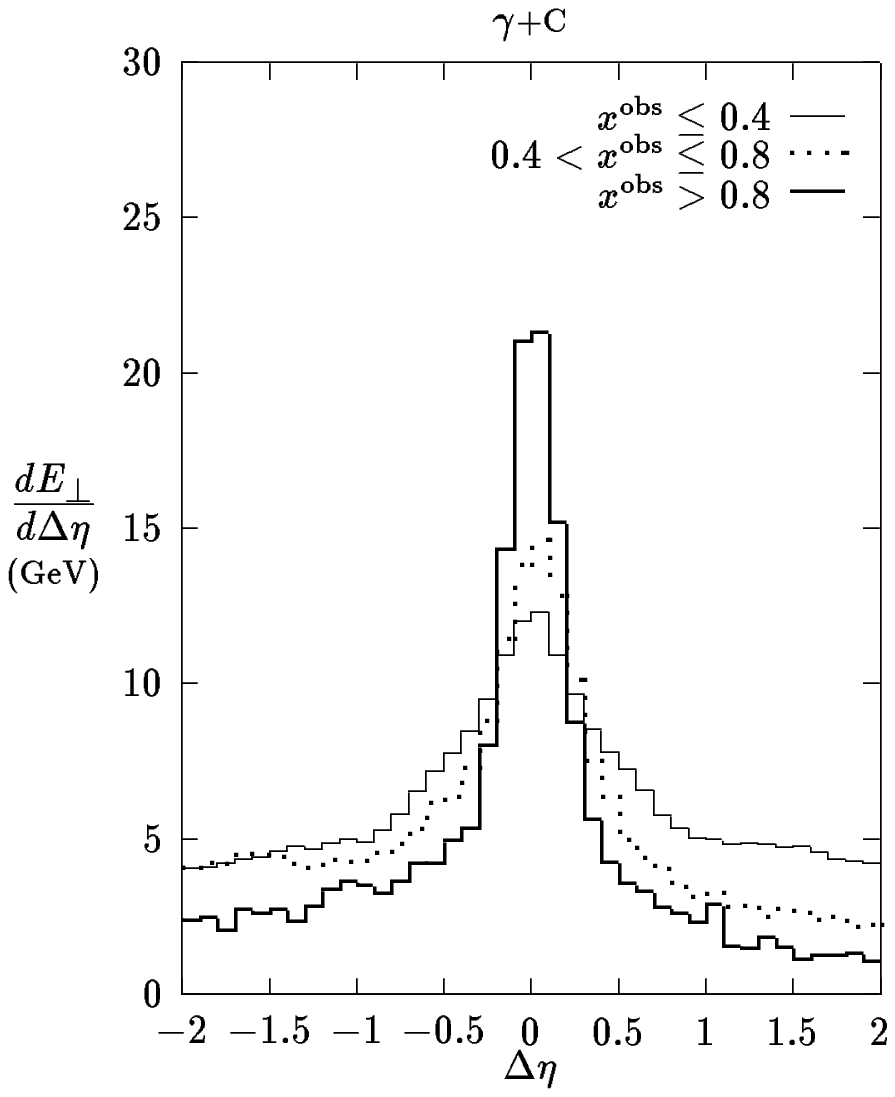}}
\put(1.0,1.1){\bf\large b)}
\put(5.0,-3.9){\psfig{figure=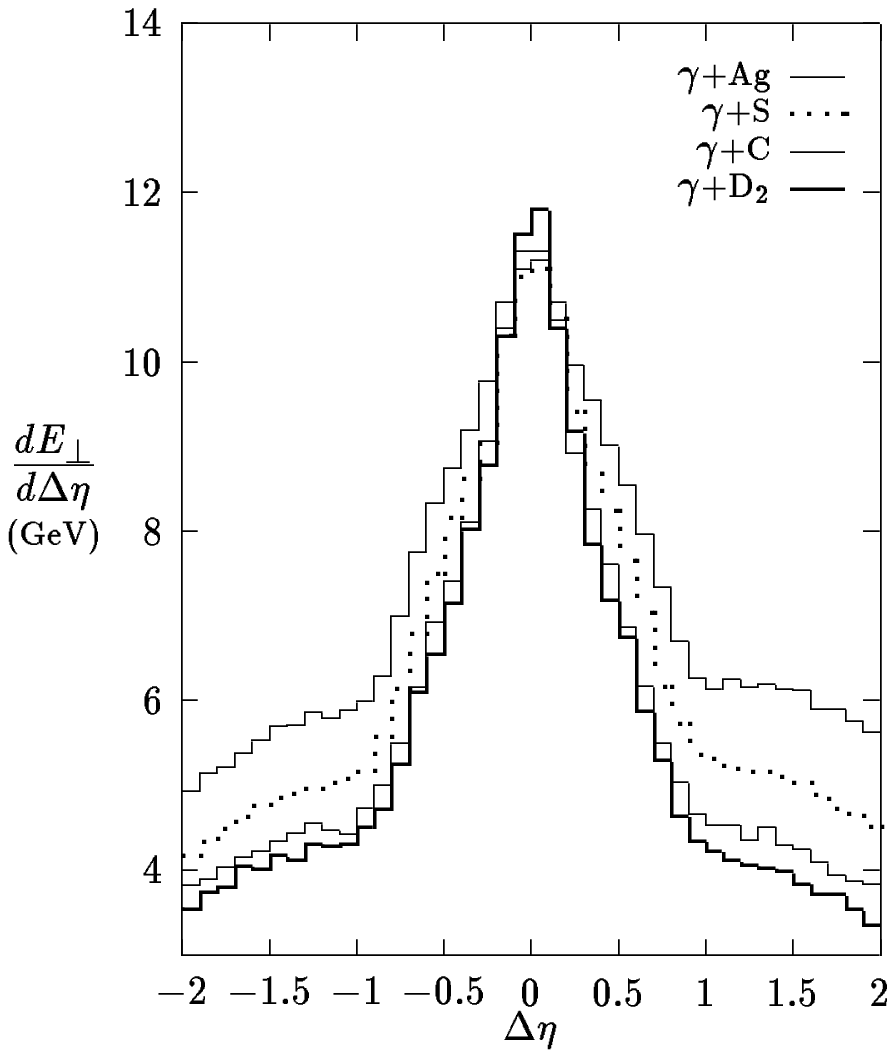}}
\put(11.0,1.1){\bf\large c)}
\put(7.5,0.0){\bf\large Fig.~\ref{gAjtxg}}
\end{picture}
\end{figure}
\end{document}